\documentclass[sigconf]{acmart}
\AtBeginDocument{%
  }

\setcopyright{acmlicensed}
\copyrightyear{2026}
\acmYear{2026}
\acmDOI{XXXXXXX.XXXXXXX}
\acmConference[CHI '26]{Make sure to enter the correct
  conference title from your rights confirmation email}{April 13--17,
  2026}{Barcelona, Spain}
\acmISBN{978-1-4503-XXXX-X/2018/06}

\usepackage{gensymb}
\usepackage{placeins}
\usepackage{caption}
\usepackage{etoolbox}
\usepackage{framed} 
\usepackage{enumitem}
\usepackage{graphicx}
\usepackage{tabularx}
\usepackage{makecell}

\begin{document}

\title{Storycaster: An AI System for Immersive Room-Based Storytelling}

\author{Naisha Agarwal}
\orcid{0009-0007-6512-1590}
\affiliation{%
  \institution{University of California, Los Angeles}
  \country{USA}
}
\email{naishaa@g.ucla.edu}
\authornote{This work was done during an internship at Microsoft Research.}
% say work was done during internship at microsoft research

\author{Judith Amores}
\orcid{0000-0003-1285-6909}
\affiliation{%
  \institution{Microsoft Research}
  \country{USA}
}
\email{judithamores@microsoft.com}

\author{Andrew D. Wilson}
\orcid{}
\affiliation{%
  \institution{Microsoft Research}
  \country{USA}
}
\email{awilson@microsoft.com}

\renewcommand{\shortauthors}{Agarwal et al.}

\begin{abstract}
 While Cave Automatic Virtual Environment (CAVE) systems have long enabled room-scale virtual reality and various kinds of interactivity, their content has largely remained predetermined. We present \textit{Storycaster}, a generative AI CAVE system that transforms physical rooms into responsive storytelling environments. Unlike headset-based VR, \textit{Storycaster} preserves spatial awareness, using live camera feeds to augment the walls with cylindrical projections, allowing users to create worlds that blend with their physical surroundings. Additionally, our system enables object-level editing, where physical items in the room can be transformed to their virtual counterparts in a story. A narrator agent guides participants, enabling them to co-create stories that evolve in response to voice commands, with each scene enhanced by generated ambient audio, dialogue, and imagery. Participants in our study ($n=13$) found the system highly immersive and engaging, identifying the narrator and audio as the most impactful elements, while also highlighting areas of improvement in latency and image resolution.
\end{abstract}





\begin{teaserfigure}
  \includegraphics[trim=0pt 300pt 0pt 0pt, clip, width=\textwidth]{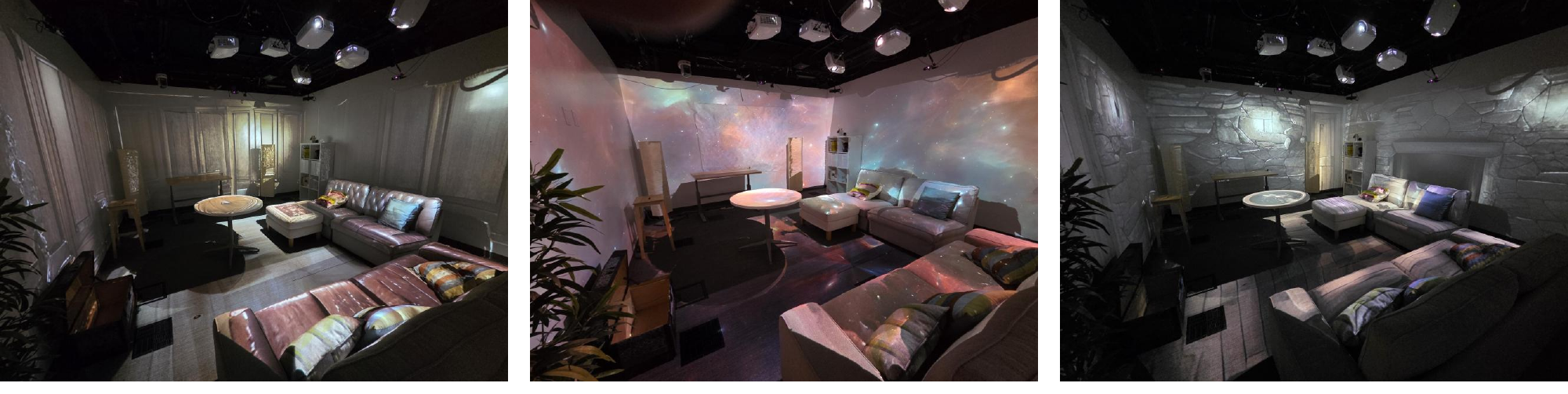}
  \caption{\textit{Storycaster} transforms your room into a responsive storytelling environment. Multiple cameras and projectors are used to augment the walls and furniture of the room, setting the scene for a story that is told through dialogue, ambient sound effects, and music. Here we see three examples of automatically generated scenes: a wood-paneled room with overstuffed leather chairs, a fantasy-inspired view into space, and a witch's cottage. Depth sensing and projection mapping are used to adapt visual content to the room’s geometry and surfaces.}
  \label{fig:teaser}
\end{teaserfigure}


\maketitle

\section{Introduction}
Spatial Augmented Reality (SAR) systems use precise geometric models of physical environments to seamlessly blend rendered virtual spaces into the real world. Systems such as Room Alive~\cite{jones2014roomalive} demonstrate how projection mapping techniques can transform a room without needing head-mounted displays (HMDs), allowing users to remain present in and directly interact with their surroundings. However, these kinds of experiences have traditionally relied on pre-authored content that cannot adapt dynamically to the user's physical environment or creative goals. Authoring such experiences is difficult since designers rarely have access to the user's room, yet convincing projection mapping requires content that is highly situated and spatially aligned with its environment. 

Recently developed AI image generation techniques offer a new opportunity to address this challenge. With the advent of large language models, we can consider the automatic generation of not just the visuals but rather the entire interaction itself, from different environments, characters, and narrative structure, all responding directly to user's creative requests. Combined with reliable speech recognition, this makes it possible for users to describe the experience they want, much as the characters in Star Trek: \textit{The Next Generation} interacted with the Holodeck.

Despite advances in SAR and immersive projection, existing systems remain limited by static, pre-authored content and require specialized technical expertise to create or adapt experiences. There is a clear need for tools that empower everyday users to author dynamic, personalized stories in their own physical environments without the barriers of head-mounted displays, complex setup, or professional design skills. 

Our goal with \textit{Storycaster} is to explore how \textbf{everyday users}, such as parents, educators, and casual storytellers, might author immersive room-scale stories through natural, conversational interactions. While professional creators have traditionally authored immersive experiences, \textit{Storycaster} lowers the barrier for non-experts, to enable the  creation of room-scale stories through intuitive, natural interaction. We envision a future where everyday environments (living rooms, classrooms, hospital rooms), become mediums for personalized creative expression, where anyone can generate these experiences on demand without technical expertise. To clarify user roles, we distinguish between (1) \textit{Storyteller}: the participant(s) who speak to the system, issue commands, and shape the narrative, (2) the \textit{AI Narrator}: which guides the story, generates dialogue, and responds to user input, acting as a co-storyteller, and (3) the \textit{Audience}: Anyone present in the room who experiences the story; the storyteller and AI narrator are both part of the storytelling process, and the human storyteller can also be a part of the audience.

This user-centered framing led several key design choices. First, we chose to use voice as the primary interaction modality because it is intuitive and aligns with the natural, conversational nature of storytelling. Voice input allows users to shape their narrative (and any changes they make along the way) without relying on additional devices or screens. Second, \textit{Storycaster} maintains users' awareness of their physical surroundings by augmenting the real environment, rather than replacing it. Unlike traditional virtual reality systems, which often immerse users in fully synthetic worlds, \textit{Storycaster} enables users to remain present and engaged in their everyday environment, and have the ability, with a simple voice command, to change their everyday surroundings. Third, we target open-ended improvisational storytelling rather than pre-structured, choose your own adventure narratives. While the system does offer branching choices, its primary aim is not to guide users through a pre-determined narrative, but rather help them invent their own through changing settings, transforming objects, introducing characters, and redirecting the plot. 

%Imagine a future of personalized entertainment, where someone can walk into any room and transform it into what they want to see - a beach in Hawaii, the top of the Himalayas, winning an award at the Oscars. Instead of having experiences and places pre-selected for people, people can choose them, in \textit{real-time}. A living room of a house turns into the Amazon rain forest, a children's hospital room transforms into the enchanting world of  Hogwarts, and a dull office station becomes the relaxing sand beaches of Cabo.

% \textit{Storycaster} is a generative AI and SAR system designed to let users author their own experiences, augmenting physical spaces with virtual worlds. It is focused on immersive storytelling and providing users with the platform to tell any story (existing or new) and see it come to life in their own room. \textit{Storycaster} distinguishes itself from virtual reality applications as its users do not lose sight of where they are and their physical surroundings. We want users to be aware of where they are, and have the ability, with a simple voice command, to change their everyday surroundings.

Previous immersive narrative experiences rely on pre-authored content. For example, the KidsRoom transformed a physical space into a story inspired by \textit{Where the Wild Things Are} through carefully designed visuals and interactions with physical props (e.g., ``now pretend the bed is a boat!'')~\cite{10.1162/105474699566297}. While groundbreaking in narrative immersion and interaction, the experience followed an entirely predetermined storyline, guiding kids through scripted interactions rather than giving them freedom to change the storyline itself. 

By leveraging generative AI, \textit{Storycaster} enables content to be created on demand, allowing environments, characters, and narratives to evolve in direct response to the audience. Through voice command, users can change environments, reimagine physical objects in the room, and choose the storyline they want to be immersed all in real-time (this includes the generation of ambient sounds, dialogue, and visual projections). In doing so, \textit{Storycaster} creates a uniquely participatory platform, where users do not just consume a story but actively co-author it. A key question is whether generative AI techniques can deliver an experience as compelling as more predetermined and hand-authored experiences such as the KidsRoom.

\textit{Storycaster}, through an orchestration of numerous GPT 4.1 driven components, designs a unique experience for every story and every room. It guides users in creating a three act narrative, using live camera feeds of the room to produce cylindrical images that are mapped back onto the walls and objects in the room. In particular, we make the following contributions:
\begin{enumerate}

    \item {\textbf{Room Scale Audio-Visual Generation}}: We present an image generation pipeline that uses generative AI techniques to create projection mapping effects that alter the look of objects in the room or create entirely new scenes. The pipeline combines SDXL with Depth Control Net and a cylindrical LoRA, taking as input live depth images of the room, and a prompt to produce cylindrical images that are projected in the room. Figure~\ref{fig:teaser} shows some example generations. Additionally we use Stability Audio 1.0 ~\cite{stability2024stableaudioopen} to create matching ambient soundscapes, rendered by a Dolby spatial audio system to position audio dynamically throughout the room.

    \item{\textbf{Narrator Agent}}: We developed a Narrator agent that guides participants throughout the experience, enabling users to co-create stories in the room. Through simple voice commands picked up by a microphone in the space, users can interact with the agent, crafting a storyline with clear actions, setting, and dialogue. The agent orchestrates the experience by generating appropriate visuals, ambient audio, and different character voices for every scene.

    \item{\textbf{Object Editing}}: We demonstrate the ability to edit at an object-level in the room, giving users the granular control to virtually augment or transform physical objects in context of an evolving narrative. We investigate the use of Grounded SAM on cylindrical images to segment out objects in the room, coupled with SD 1.5 inpainting to generate virtual overlays and map them on corresponding objects in the room.

    \item{\textbf{Evaluation of \textit{Storycaster}}}: We conducted a study with thirteen participants and three pilot sessions. Each user was able to create one or more stories with \textit{Storycaster} in a 20-25 minute experience. Through a pre and post experience questionnaire and a semi-structured interview, we evaluated \textit{Storycaster} in its ability to immerse users in a story, as well as determine which part(s) of the experience contributed most to user's overall enjoyment. Our findings offer insights for future open-ended immersive storytelling experiences, and their broader applications.

\end{enumerate}

\section{Related Work}

\subsection{CAVE Systems}
Cave Automatic Virtual Environment (CAVE) systems \cite{cruz-neira1993cave} are projection-based systems that are designed to immerse virtual reality in a physical space. From CRT projectors to multi-projector CAVE systems with commodity hardware~\cite{andujar2005multiprojector} to the StarCAVE~\cite{DEFANTI2009169}, they have evolved over the years, adapting through technological changes. Most recently, CAVE systems have extended beyond projectors, replacing projection screens with flat panel displays ~\cite{article}. In all cases, they are effective in having immersive room-scale VR environments that blend together virtual imagery with the user's physical space. However, there has been limited work, to the authors' knowledge, on CAVE systems where users can use generative AI techniques to deliver custom narratives and visuals in their environments. This is what \textit{Storycaster} aims to address.

\subsection{Immersive Storytelling}

Immersive storytelling departs from traditional, linear, single-medium narratives by inviting audiences into fragmented stories that unfold across multiple forms of media ~\cite{polydorou2024immersive}. Within the Metaverse, a digital environment enabled by virtual reality (VR), storytelling takes shape through four distinct dimensions— agency, interactivity, immersion, and community— that distinguish it from conventional approaches ~\cite{gantayat2024traditional}. These qualities foster personalization and a sense of co-authorship, as each user actively shapes their experience. VR and mixed reality further expand the possibilities of narrative by enabling audiences to inhabit diverse scenarios and perspectives in both real and fictional worlds ~\cite{doyle2017immersive}. Projects such as RealityMedia demonstrate how VR can function as a testbed for narrative experimentation, where spatial context, collections, and artifacts become central to story construction in ways unavailable to traditional media ~\cite{realitymedia}. Similarly, ConnectVR explores audience-driven storytelling through a trigger–action framework, allowing characters and objects to adapt their behaviors in response to user interaction ~\cite{Chen_2024}. In the realm of audio, there has been work on creating personalized soundscapes and immersive audiobooks ~\cite{sonora, selvamani2025multiagentaiframeworkimmersive}. Despite these advances, a common thread persists: immersive narratives still depend on physical devices—such as headsets or mobile phones—for participation. Little work has yet explored how to craft immersive stories that unfold directly within physical space, without the mediation of external devices.

\subsection{Narrative Creation with Generative AI}

Generative AI has recently become a powerful tool for storytelling ~\cite{it2025integrating}, enabling audiences to engage in back-and-forth conversations with AI narrators ~\cite{caramiaux2025generativeaicreativework}. Such personalized narratives not only foster creativity but can also help individuals reflect on and process their own experiences and emotions ~\cite{Fabre_2025}. To support this, systems like the Universal Narrative Model ~\cite{gerba2025narrativecontextprotocolopensource} propose standards that leverage the strengths of generative AI while preserving the author’s intent and creative vision.

A variety of frameworks have been developed to explore narrative generation, including approaches to branching storylines ~\cite{huang2024whatifexploringbranchingnarratives, yang-etal-2023-doc, yang2022re3generatinglongerstories}, different kinds of narratives including movie scripts ~\cite{zhu2022leveraging}, and the design of novel storyworlds ~\cite{johnson-bey2023toward, 10.1145/3586183.3606763}. Tools such as StoryForge ~\cite{storyforge} provide user-friendly interfaces that balance human authorship with AI collaboration, while projects like ID-8 ~\cite{antony2024id8cocreatingvisualstories} integrate visual and audio elements to create rich, multi-modal experiences. Beyond written narratives, AI has also been a transformative force in interactive games ~\cite{santiago2023rollingdiceimagininggenerative}, extending from early text-based adventures like Zork ~\cite{1658697} to collaborative role-playing systems such as PANGea ~\cite{Buongiorno_Klinkert_Zhuang_Chawla_Clark_2024} and 1001 Nights ~\cite{Sun_Li_Fang_Lee_Asadipour_2023}.

\subsection{Uniting Physical and Virtual Worlds}

Immersion lies on a spectrum—from augmented to virtual to mixed realities—where tangible interaction remains a central tenet \cite{spittle2022interaction}. Inspired by Ishii’s Tangible Bits philosophy, the goal is seamless integration between digital content and physical experience~\cite{ishii}. This is seen in early work with toys and augmenting these into children's storytelling ~\cite{glos1997once}. 

Several systems generate VR environments based on real-world physical structures. VRoamer creates VR scenes that dynamically align with unknown building layouts, enabling safe continuous walking ~\cite{cheng2019vroamer}. DreamWalker substitutes real-world walking with virtual cityscapes, overlaying virtual assets onto physical paths ~\cite{yang2019dreamwalker}. Sra et al. ~\cite{sra2016procedural} procedurally generates VR environments from reconstructed indoor spaces. All of these systems leverage physical geometry to inform environment substitution within VR, maintaining safety during movement and enabling large scale-exploration.

While \textit{Storycaster} shares the idea of grounding virtual content in physical geometry, it diverges in several ways: (1) \textit{Storycaster} focuses on storytelling with no headsets, where users remain visually aware of the room they are in, (2) VRoamer/DreamWalker replace the environment, while \textit{Storycaster} enables users to author environments, characters, and narratives through voice commands, and (3) \textit{Storycaster} includes more granular control including object-level editing to turn physical objects into narrative elements (an interaction not present in these VR systems).

Other recent AI+XR systems, such as LLMR~\cite{llmr}, ImaginateAR~\cite{lee2025imaginatearaiassistedinsituauthoring}, DreamCodeVR~\cite{dreamcodevr}, Dreamcrafter~\cite{dreamcrafter}, and VRCopilot~\cite{vrcopilot}, have explored scene authoring and editing in extended reality, often via natural language interfaces. However, these typically require external screens or head-mounted displays, and do not offer the in-situ, room-scale immersion that \textit{Storycaster} provides. While some storytelling systems are beginning to integrate more object-centered interactions in order to shape narratives in physical space~\cite{ocadu4471}, these still lack the 360\degree~immersion that a room provides. Work such as IllumiRoom ~\cite{jones2013illumiroom} takes a step in this direction, creating immersive experiences for gaming on TVs, turning the surroundings of the device into the environment of the game itself. Similarly, Room Alive provides a framework for turning any room into an immersive gaming experience ~\cite{jones2014roomalive}.

\textit{Storycaster} is a novel step beyond these: it enables users to construct narrative worlds using simple voice commands, while spatial reasoning grounds the experience directly within the physical space. In doing so, \textit{Storycaster} bridges the immersive room-scale presence of CAVE systems, the narrative flexibility and co-creative potential of modern immersive storytelling, the generative power of AI-driven narrative construction, and the seamless physical and digital world integration.

By combining these threads, \textit{Storycaster} seeks to redefine immersive storytelling as a physically embodied dynamically controllable narrative experience, rooted in the user's real environment itself.

\begin{figure*}
    \centering
    \includegraphics[
    trim=0pt 150pt 100pt 100pt, %left bottom right top
    clip,
    width=5in          
  ]{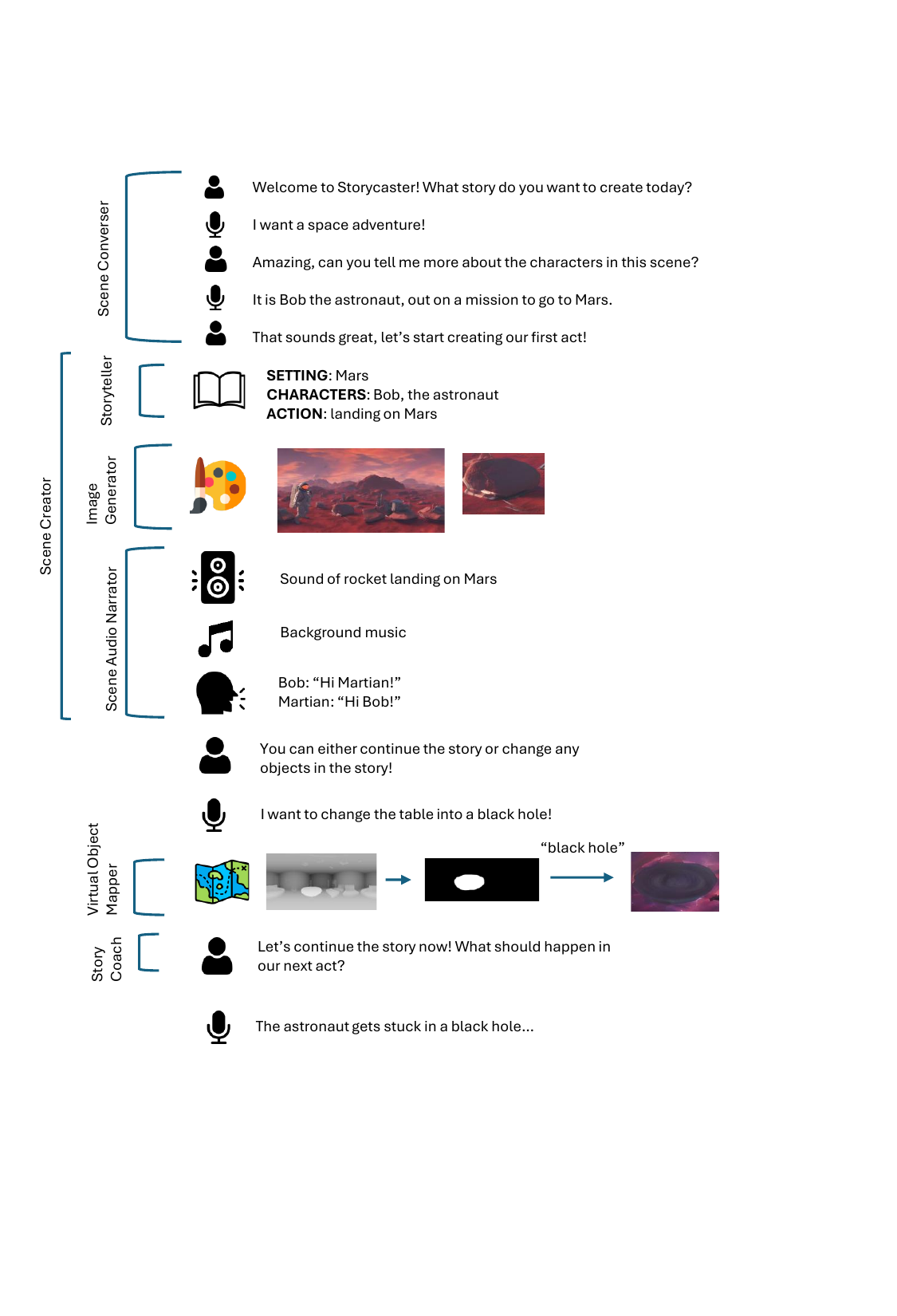}
    \caption{System overview of \textit{Storycaster}. The Narrator orchestrates the storytelling process, prompting the user for input and guiding them through three acts. The Scene Converser gathers initial story elements, which are passed to the Scene Creator. Within the Scene Creator, the Storyteller generates narrative text, the Image Generator creates visuals, the Virtual Object Mapper overlays virtual counterparts onto physical objects via inpainting, and the Scene Audio Narrator produces ambient audio and character dialogue. The Story Coach manages transitions between acts, allowing users to either continue the narrative or change objects in the room. Segmentation, inpainting, ambient audio generation, and image generation all happen on the server side. This process repeats for three acts, after which the user can choose to begin another story or conclude the session.}
    \label{fig:narratorarch}
\end{figure*}

\section{Storycaster}
\textit{Storycaster} brings together various integrated components to create an immersive storytelling experience. The system includes: (1) an image generation pipeline that generates 360\degree{} cylindrical images from live camera feeds and maps them back on the walls and objects in the room, (2) a Narrator Agent that guides participants throughout the experience, and (3) object-editing capabilities that allow users to transform physical objects into virtual overlays contextualized within an overarching story. The overall architecture of the system is described in Figure \ref{fig:narratorarch}, and the prompts used by each module are discussed in the Appendix.

When first entering the room, users are welcomed into a projected library environment, accompanied with gentle background music. Bookshelves line the walls, encouraging users to ``choose'' a book from the shelf or make a story of their own. They are then guided through a tutorial section that demonstrates the room's capabilities: image generation, ambient audio generation, and object-level editing. Participants interact by speaking aloud; their input is captured by a room microphone, transcribed with Azure Speech to Text services, and processed by the Narrator agent. A chime sound signals after a user's input has been registered, reassuring them that their voice has been heard. Additionally, users are asked if they have brought in a personal object; this object is later integrated into the stories they create, acting as a ``Macguffin'' in the story, a central talisman for which the story revolves around ~\cite{brewer1992twentieth}. Similar work was done in ``Rosebud'' that treated a child's toy as a central object in a story they created \cite{glos1997rosebud}.

Once participants are comfortable with the system, they transition into the actual story experience. The experience follows a three-act structure commonly used in film: The Setup, The Confrontation, and the Resolution \cite{field1979screenplay}. Users begin by engaging in a conversation with the Narrator that helps them craft the initial premise for their story, the agent probing for characters, the setting, and the main action taking place to situate the story. Once the narrator has enough information to create the first act, it asks the user for confirmation, and proceeds, generating the script, visuals, ambient audio, and the different voices for characters in the scene.

At the end of each act, users are prompted to either continue the story or modify elements in the room. If they choose to continue the story, the Narrator presents three possible directions for the narrative, while also allowing users to provide their own storyline or let the Narrator decide. This repeats for all three acts of the story, culminating in users creating a new story or ending the session.

\subsection{\textbf{Room Features}}

For sensing and display there are six projectors and four Azure Kinect cameras mounted in the ceiling of the room. The cameras take in both color and depth images. The cameras and projectors are mounted to roughly cover the full room below the ceiling and are calibrated as in~\cite{jones2014roomalive}. 
Additionally, there is a Dolby Atmos spatial audio system set up for sound playback as well as a microphone for user input.

The current implementation was installed in a single-calibrated projection-mapped room; however, the underlying approach is designed to be adaptive. \textit{Storycaster} can theoretically operate in any room equipped with calibrated projectors and depth cameras, since the pipeline does not rely on room-specific characteristics or dimensions. However, environmental constraints such as cluttered spaces, or non-rectangular rooms can reduce projection quality. Future work will explore a more portable lightweight version of our setup.

\begin{figure*}[h]
  \centering
  \includegraphics[trim=0pt 100pt 0pt 0pt, clip, width=\linewidth]{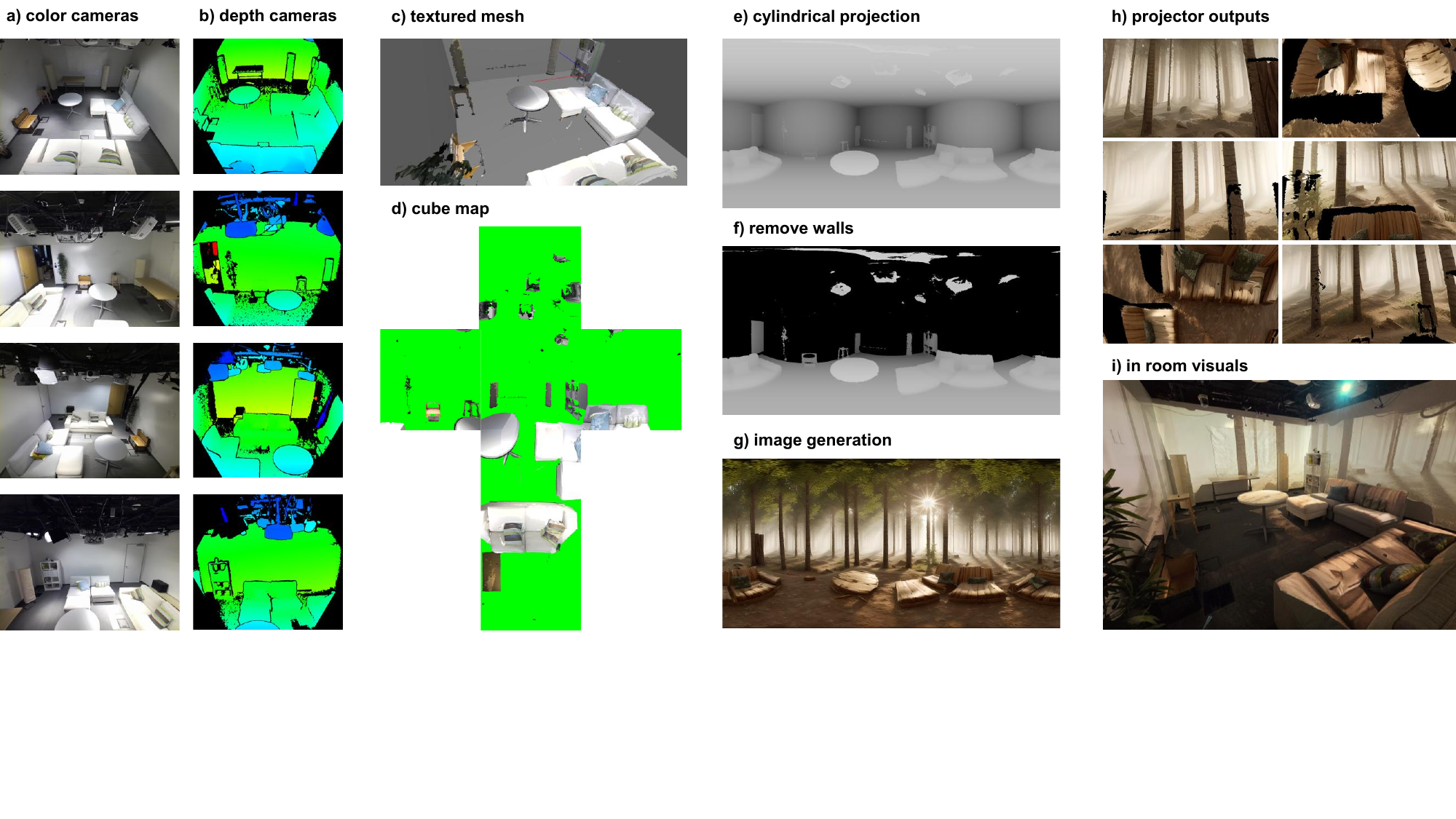}
  \caption{\textit{Storycaster} combines multiple color and depth cameras to generate visuals that follow the physical features of the room. Calibrated color (a) and depth cameras (b) (rendered as false color) are combined to create a textured mesh that serves as a realtime geometric model of the room (c). This model is rendered to a single cube map (d). The cube map z-buffer is re-rendered as a cylindrical (or 360\degree) depth image (e). To render an exterior or outdoor scene, walls may be optionally removed (f). Stable diffusion is applied with a depth ControlNet conditioned on (f) and a 360\degree~LoRA model. ControlNet effects are masked (eliminated) for pixels that are zero in (f), allowing for a more natural horizon effect in exterior (outdoor) settings. The generated image (g) is upscaled 4x (not shown). The image displayed at each projector is calculated from the geometric model of the room, using intrinsic and extrinsic parameters of the projectors (h). Black pixels correspond to regions of the geometry that are not visible in (e) (i.e., they are occluded by another object). The final projected visuals create 360\degree~views into an exterior virtual space and change the appearance of objects in the room (i).
}
  \label{fig:image_generation}
\end{figure*}

\subsubsection{\textbf{Image Generation}}

The projectors display imagery that can subtly alter the look of the room or more dramatically transport the viewer to an altogether different space. These effects are created to follow and support the narrative, using generative AI techniques such as stable diffusion.

To generate the images, a cylindrical (360\degree equirectangular projection) depth image of the room is first assembled from the multiple live depth camera images. We opted for cylindrical images as this allows for a one-shot approach of generating an image that can be projected throughout the entire room, rather than generating multiple separate views. A depth ControlNet uses the cylindrical depth image as guidance to encourage image generation to follow the physical structure of the room, including the walls and furniture.

We utilize two models for image generation: SDXL for generating full image scenes and SD 1.5 for object-level inpainting. In both cases, the live cylindrical depth image of the room and the prompt is passed into the model which is augmented with a depth ControlNet~\cite{zhang2023adding}, and LoRA tuned to generate cylindrical images. Through experimentation, we used a value of 0.76 for the depth ControlNet strength value for SDXL, and 0.54 for inpainting with SD 1.5 (refer to Appendix \ref{fig:appendix-depth-sdxl} for these results). This produces 360\degree~cylindrical images that are then upscaled and projected back onto the four walls of the room, using projection mapping techniques. The image generation process is illustrated in Figure~\ref{fig:image_generation}.

While the depth control net ensures that wall surfaces and room geometry influences image placement and perspective, the semantic meaning of physical objects is not explicitly modeled during scene-wide image synthesis, unless explicitly invoking the object-editing interaction. The prompts rather focus on describing the setting of the scene (``deep forest at dusk'', ``library'') and letting the model take creative liberties with what each object becomes. This design choice allowed generations to remain robust across diverse stories but also resulted in mixed user feedback regarding object-level coherence. We discuss this further in Section ~\ref{sec:design}. When users initiate object-level editing (e.g., ``change the table into a black hole''), the system incorporates the semantic label of the physical object into the inpainting prompt, enabling targeted transformation. This approach allows users to personalize their stories by re-purposing familiar items as narrative anchors. Our user study found that object-level editing enhanced personalization and narrative engagement, as participants enjoyed transforming real objects into story elements (see Section ~\ref{sec:results}).

\subsubsection{\textbf{Audio}}

For generating ambient audio, we utilize Stability Audio 1.0, with outputs designed to loop continuously for seamless playback.

Additionally, we define an \texttt{AudioPlayback} API that manages multiple sound sources in the room, each defined with a unique name, a volume, and spatial coordinates. The API allows multiple sources to be added, replayed, or removed. Each sound source is instantiated as a separate connection, and played sequentially through the spatial audio system.

\subsubsection{\textbf{MCP Server}}

In order to seamlessly integrate the LLM driven storytelling components with the room, we implement a Model Context Protocol (MCP) architecture for our server-client setup. The server exposes several APIs for image and audio generation, allowing the client to orchestrate story structure while offloading computation-intensive tasks to the server.

\subsection{The Narrator Agent}
The Narrator agent functions as a main orchestrator, managing multiple LLM-driven (GPT 4.1) components. These include a Scene Converser, and multiple instances of a  Scene Creator, each consisting of a Storyteller, Image Generator, Scene Audio Narrator, a Virtual Object Mapper, and a Story Coach. The overall architecture and module interaction are illustrated in Figure \ref{fig:narratorarch}.

The Narrator includes a \texttt{speak} function that is called for each piece of text generated. This function takes in the text to be recited, the type of voice, and mode, which varies based on whether the system is in tutorial or storytelling mode. Each mode utilizes a different style prompt: the tutorial mode uses a more relaxing calm voice, while storytelling mode uses an engaging, friendly delivery. This style prompt and the text is passed into GPT's text to speech model (TTS), which outputs audio at 24 kHz. The audio is resampled to 48 kHz, and sent to the AudioPlayback API to be played through the room's speakers. We chose to use the TTS model instead of the GPT Realtime model, despite its more natural conversation flow, due to the level of control and structure needed for the experience. 

When the Narrator prompts users for input, they can simply speak in the room. A microphone captures their voice, which is streamed via a websocket URL to Azure Speech-to-Text services for transcription. To handle pauses or hesitations without cutting off transcriptions prematurely, we increase the end-of-segmentation silence timeout to 3 seconds. Once the speech recognizer determines the user has finished speaking, it returns the final transcription, closes the stream, and plays a chime sound to signal the Narrator has successfully received the input.

% system performance and latency
We recorded performance metrics to characterize system responsiveness. It took approximately 7.3 seconds to produce images (both scene-level and object-level) including upscaling, and 12 seconds for ambient audio generation. During the image generation, users were able to see the images form in the room from the initial model generation to the eventual final version after upscaling. In terms of generations, during our user study, all generations were generally successful and correspondent to user's requests determined by their responses to an interview (see Section ~\ref{sec:results} for details). Future work includes reducing these latency values so the experience is as seamless and smooth as possible.

Each module of the Narrator agent is described in detail in the following sections:

\subsubsection{\textbf{Scene Converser}}
When first starting the story creation process, the Scene Converser asks the user what kind of story they would like to create. It engages in a back and forth conversation with the user to craft an initial premise for their tale, probing for four key elements: (1) the setting, (2) characters, (3) main action, and (4) the mood and tone of the scene. The Converser is designed to keep the instruction lightweight, avoiding excessive questioning as details can be refined later in the story. 

After each user response, the Scene Converser evaluates whether it has enough information to proceed creating the scene. This is determined by both the information the user provides, as well as the number of conversational turns. If yes, then it will respond with ``READY''; otherwise, it responds with ``NEED MORE''. Once ``READY'' is returned, the Scene Converser crafts a personalized prompt asking the user if they are ready to proceed with bringing their story to life. It then analyzes the user's response to identify if they have agreed or want to continue the story (such as ``yes'', ``I am ready'', ``let's go''). If it determines positive confirmation, it creates an instance of a Scene Creator and starts the scene creation process. If the user wishes to add more details, the Converser continues conversing until the user is satisfied. For users who already know exactly what story they want to tell, they can skip the back-and-forth with the Converser entirely by indicating readiness, after which the Converser confirms, and proceeds creating the scene.

\subsubsection{\textbf{Scene Creator}}
Once a user is ready to create a scene, an instance of a Scene Creator is created. The Scene Creator is responsible for three components: the Storyteller, in charge of the story text; the Image Generator, in charge of image generation and object editing capabilities; and the Scene Audio Narrator, in charge of generating both ambient soundscapes and character voices. Together, these classes transform user's input into a multimodal storytelling experience.

\subsubsection{\textbf{Storyteller}}
The Storyteller is in charge of generating the actual story script. It takes in the following inputs: (1) the \textbf{current act} (first, second, or third), (2) any \textbf{previous scenes} generated, (3) the \textbf{original storyline} from the user's initial interaction with the Scene Converser, (4) the user's \textbf{personal object} description from the tutorial, and (5) the current \textbf{user input}. The original storyline and the current user input are the same if creating the first act. 

Using these inputs, a detailed system prompt is crafted to generate a scene based on these personal preferences. In order to make the stories feel more like a play with heavy dialogue, we use few-shot prompting of various movie scripts from the IMSDB movie script database ~\cite{imdb}. We hand-selected four such scripts from this database, and extracted out the first 25 lines of two random scripts for each scene generation, ensuring variety across story structures. In particular, we aimed to emphasize \texttt{setting}, \texttt{dialogue}, and \texttt{actions} in the generated stories---something that movie scripts do best. Guided by these examples and detailed instructions on key storytelling principles---such as prioritizing character dialogue, using concise visual descriptions, and revealing backstory and motivations behind character's actions---the Storyteller generates a 8-10 sentence scene.

\subsubsection{\textbf{Image Generator}}

Once the Storyteller generates the current scene text, it is passed to the Image Generator, which is responsible for both the wall projections and object-level generations in the room.

For room-scale visuals, the scene text is used to generate an appropriate prompt for SDXL, the image generation model. The system prompt for this conversion was carefully crafted, and includes few shot examples pairing different story excerpts with corresponding  visual prompts of no more than 50 words. Additionally, based on insights from pilot studies, we added explicit instructions to favor ``saturated bright colors'' as we observed this improved the vibrancy of the generated images in the room.

The generated prompt is also used to classify if there should be a removal of walls in the generation. This was determined by classifying the specified scene as outdoors or indoors; outdoor environments would enable the removal of walls and indoor environments would not. Enabling the removal of walls for outdoor environments allowed for visuals to extend beyond the four walls and create a sense of openness. Indoor environments, by contrast, preserve wall constraints to maintain spatial consistency.

Finally, the refined prompt, classification output, and a white mask (ensuring the full depth image of the room is used) are passed into the MCP image tool to generate the final projection.

\subsubsection{\textbf{Virtual Object Mapper}}

The Virtual Object Mapper, a component of the Image Generator, is responsible for generating appropriate virtual object mappings to the physical objects in the room that align with the current story. This includes detecting objects, generating mappings between physical and virtual items, and producing inpainting prompts to pass into the server-defined inpainting tool. This process is performed for every scene, and users may also choose to remap objects of their choice between acts.

For object detection, we found that Grounded SAM did not consistently produce reliable masks (please see Section \ref{object_editing}). Instead, we opted for a more manual approach where we hand-crafted masks of various objects in the room (e.g., table, ottoman, lamp, and sofas). Each mask file is labeled in the format \texttt{mask{object}white.png}, enabling straightforward retrieval.

Once objects are identified, the system generates mappings using the following inputs: (1) The current scene, (2) the previous scene for consistency between two scenes, (3) the detected physical objects in the room, and (4) the previous mappings generated (if applicable). Using few shot prompting of different physical objects and potential virtual overlays they can be converted to, the system prompt lays out mapping guidelines that emphasize story consistency, physical compatibility with the shape of the object, and continuity between previous mappings. The output includes both a mapping of physical to virtual objects, and an optimized inpainting prompt.

Finally, for each object, its corresponding mask is retrieved and converted to a base64 string. This string, combined with the inpainting prompt, is passed into the image tool to generate overlays, effectively transforming real objects in the room into story-appropriate virtual artifacts.

\subsubsection{\textbf{Scene Audio Narrator}}
The final component of a Scene Creator instance is the Scene Audio Narrator, in charge of both the ambient audio generation and the voice acting for a scene.

To generate ambient audio, the scene narrator takes in the generated scene text from the Storyteller and using a carefully designed system prompt with few-shot examples of story-audio pairs, constructs an appropriate concise description of the desired soundscape. This prompt is then passed into the server-side ambient audio tool to generate the sounds.

Additionally, background music is played in between acts through searching Spotify~\cite{spotify_web_api} for top 5 playlists that meet the prompt ``background instrumental music'' and plays a song from the top search. This way, users do not hear the same music for each act, and in between stories.

Once the ambient sound has been created, the Scene Audio Narrator produces dialogue for the various characters in the scene, using the GPT TTS model. This process consists of (1) parsing the scene text into individual lines, each one assigned to a specific character (or the Narrator), (2) assigning a distinct vocal style to each character based on dialogue content and story context (passed in as the ``instructions'' parameter in TTS), (3) assigning a voice for each character from GPT's voice selections (alloy, echo, fable, onyx, nova, and shimmer), determined on the scene text and their voice style assignment, and (4) rendering the recitation of voices. Additionally, to reinforce spatial immersion, each character is assigned a fixed location in the room, depending on the number of the characters in the scene. This allows for conversations to play out across the room and for characters to speak from different locations. Consecutive lines by the same character are grouped together to streamline recitation.

Dialogue playback is managed by the server's \texttt{AudioPlayback} API, which takes as input the sound source name, volume, and spatial position. All TTS instances for a scene are generated simultaneously, ensuring synchronized timing, and played sequentially to avoid latency between lines. Once narration concludes, all connections are closed, finalizing the scene's audio performance.

\subsubsection{\textbf{Story Coach}}
After a scene concludes, the Narrator prompts the user to either continue the story or modify any objects in the room. If a user indicates they want to continue the story, the story transitions to the next act and the Narrator calls the Story Coach. If the user indicates they want to change an object in the room (e.g.,``I want to change the ottoman to a campfire''), their request is parsed to identify the physical object and its desired virtual counterpart. The corresponding mask is retrieved, and the inpainting API is directly called with these inputs.

For continuing the story, the Story Coach takes in the following inputs: (1) the original storyline established with the user's initial interactions with the Scene Converser, (2) how many acts have happened so far and what act we are currently at, and (3) the most recent scene text. From these, the Story Coach provides three possible directions for the narrative, tailored to the current act. Users can then: (a) select one of the provided options, (b) propose their own idea, or (c) delegate the decision to the Narrator. For option (a), an additional LLM call extracts the full text of the selected option, which is then used to generate the next act. For option (b), the user's idea is passed directly as input with no further questioning. For option (c), the Narrator generates the next part of the story, building on prior events.

\subsubsection{\textbf{Wrapping Up}}

The described process above repeats until all three acts are complete. After the third act, the Narrator congratulates users on completing their story, and asks if they would like to create another one. If yes, a new instance of the Narrator is launched; otherwise, the experience ends. 

\begin{figure*}
    \centering
    \includegraphics[width=\linewidth]{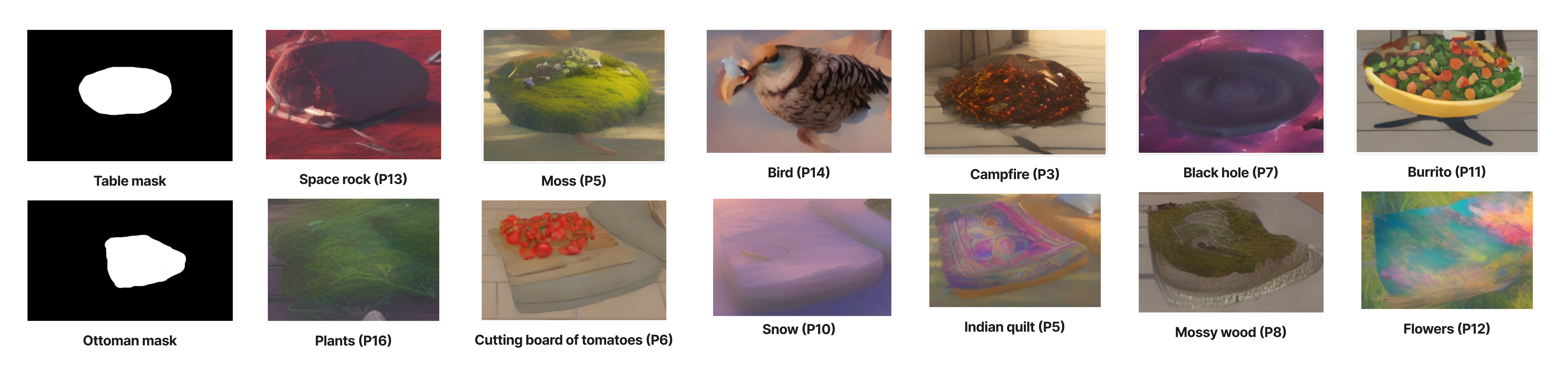}
    \caption{Examples of object editing from user studies. Participants selected real-world objects (table and ottoman, shown with their masks on the left) and reimagined them as virtual objects within their stories. The resulting generations (right) illustrate diverse mappings such as a campfire, quilt, black hole, burrito, and flowers, demonstrating the flexibility of object-level transformations in \textit{Storycaster}.}
    \label{fig:objectediting}
\end{figure*}

\subsection{Object Editing}\label{object_editing}

One of the key interactions in this experience is the ability for users to edit at an object level. This granular control allows users to select a physical object in the real world and turn it into one in the story world, while still preserving the object's underlying shape and size. See Figure \ref{fig:objectediting} for examples taken from our user study.

To detect physical objects, we apply Grounded SAM to segment out objects and their masks from a cylindrical image of the room. By using the label ``all objects,'' the model extracts masks for every visible object without requiring predefined categories. These masks, along with the room image, are then passed to GPT-4o to generate textual descriptions of the detected objects.

In practice, we find that grounded SAM performs well for objects in the central region of cylindrical images, and less so for objects on the top or bottom. For items of similar shapes and sizes (ex. ottoman and table), we found that Grounded SAM does not return consistent results, oftentimes confusing one for the other. For this reason, we hand-crafted masks for key objects in the experience so we could be sure that the object a user requests is the one being transformed.

\begin{figure*}
    \centering
    \includegraphics[width=\linewidth]{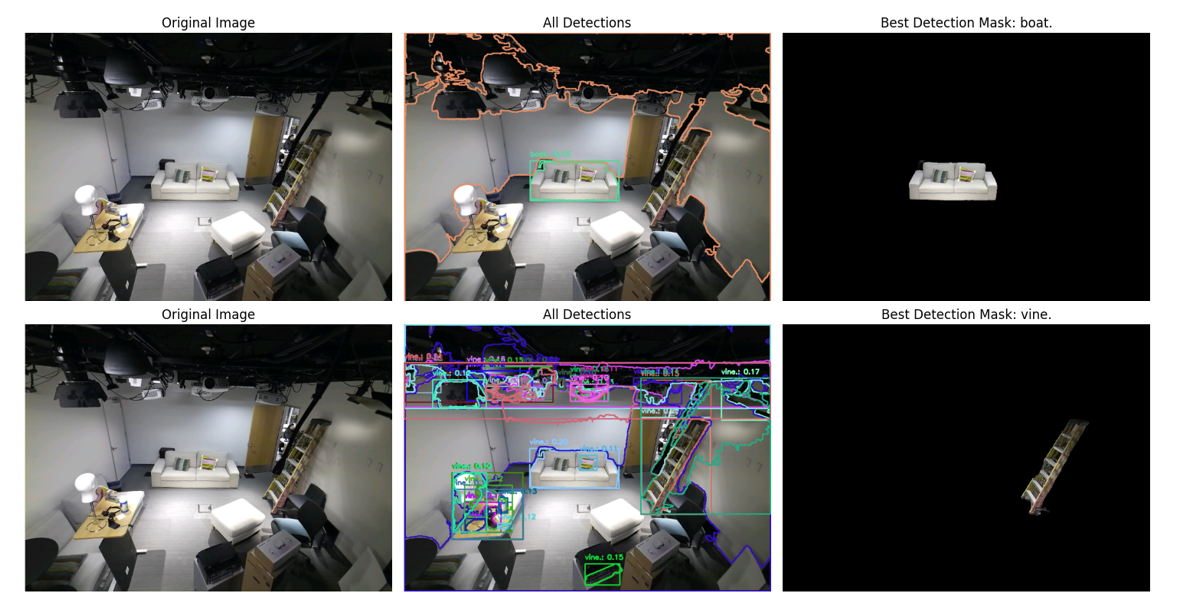}
    \caption{Examples of using Grounded SAM to identify virtual objects by passing in object prompts with a lowered similarity threshold (0.1). Top row: prompting “boat” highlights the sofa as the closest visual match. Bottom row: prompting “vine” highlights the ladder. From left to right: original room image, all detected masks, and the best detection mask. This demonstrates the ability to flexibly map real objects to virtual counterparts, enabling novel object-level editing in immersive storytelling.}
    \label{fig:sam}
\end{figure*}

Interestingly, we also discovered that grounded SAM can be used to identify \textit{virtual} objects that are not in the room itself. By passing in the name of a virtual object and lowering the similarity threshold (we used 0.1), the model correctly identifies objects in the room that look most closely to the specified one. For instance, as seen in Figure \ref{fig:sam}, prompting ``boat'' returned the mask of a sofa, while prompting ``vine'' highlighted the ladder. This capability enables flexible mappings between real and imagined worlds, and to our knowledge, is a novel contribution to the system's semantic reasoning and interpretive flexibility, particularly in the context of object-level editing.

Object-level editing was included not only to demonstrate technical feasibility but to explore a new feature to support narrative making. Unlike scene-wide visual changes, object transformations allow users to re-purpose familiar physical items (sofas, tables, lamps) as narrative anchors. Even when generative quality varied, object editing created a connection between the virtual and physical worlds, enabling interactions such as turning an ottoman into a campfire, or a sofa into a boat. This form of re-contextualization is rare in prior AI + XR systems, and is a key distinction of \textit{Storycaster}.

\captionsetup{skip=10pt}

\begin{figure*}[p]
  \centering
  \includegraphics[width=\linewidth,height=0.9\paperheight,keepaspectratio]{figures/generations_p1.pdf}
  \label{fig:grid}
\end{figure*}

\begin{figure*}%\ContinuedFloat
  \centering
  \includegraphics[width=\linewidth,height=0.9\paperheight,keepaspectratio]{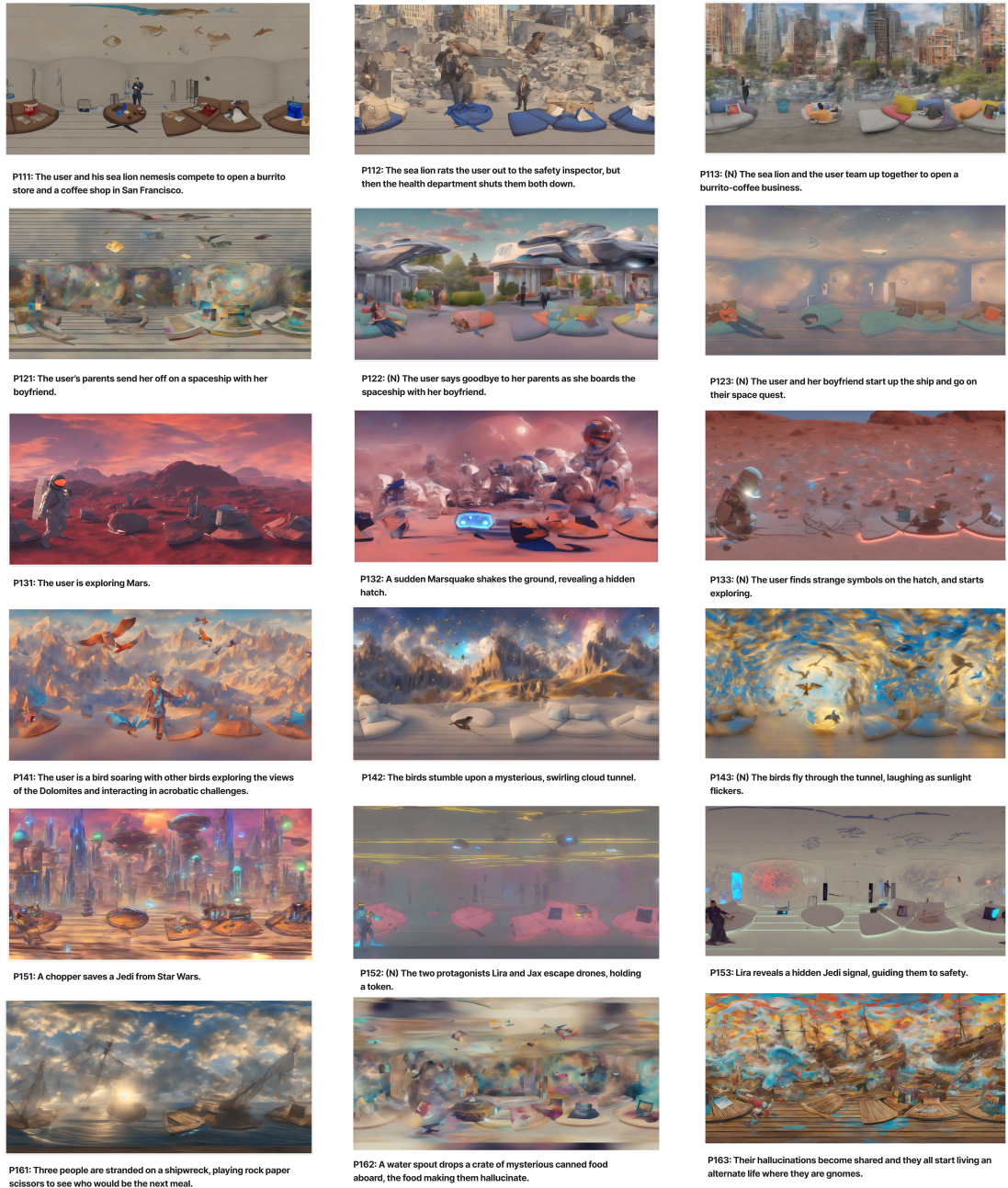}
  \caption[]{Generations from user studies ($n=13$). Each row represents one participant, with three images corresponding to the acts of their story. Labels indicate participant ID and act number (e.g., P51 = Participant 5, Act 1). Images marked with (N) denote acts where the user allowed the Narrator to decide the direction of the story. The examples highlight the diversity of user-authored narratives, ranging from historical explorations and mythological tales to domestic scenes, space adventures, and fantasy worlds, demonstrating the flexibility of \textit{Storycaster} in supporting a wide variety of storytelling themes.}
      \label{fig:generationsStudies}
\end{figure*}

\section{User Study}

We conducted a user study to explore how users felt about the experience and the type of content they would generate. In particular, we were interested in exploring (1) how immersed users felt in the experience, (2) what specific features of the experience added the most to their overall enjoyment, and (3) how users reacted to the physicality of the room while still being able to create a virtual world. Figure \ref{fig:generationsStudies} shows examples of the outputs generated by participants during the study, which we further describe in Section \ref{sec:results}. 

This evaluation was intentionally framed as a usability exploration, rather than a comprehensive benchmark against screen-based storytelling or alternate XR interfaces. Our goal was to understand how users interacted with the system, and what new forms of immersion a system like this enabled. We acknowledge the value of comparative baselines, such as 2D generative storytelling, scripted versus generative conditions, and with/without object-level editing. These comparative conditions are important for quantifying \textit{Storycaster's} unique contributions and will be the focus of future studies. In this initial work, we prioritized open-ended, in-situ interaction to capture authentic user experiences and preliminary behaviors. 

\subsection{Participants}
A total of 13 participants participated in the study, with three additional one being pilot studies. Age was collected in ranges (e.g., 25–34, 35–44). Most participants were in the 25-43 age group. To estimate the average age, the midpoint of each range was used. The resulting average age of participants was approximately 34.1 years. Out of the 13 participants, there were 5 females, 7 males, and one gender variant/non-conforming participant. 

All subjects reported interacting with AI systems on a daily level for technical tasks (coding, writing, data analysis), creative tasks (brainstorming, prototyping, generating images/video), and support tasks (summarization, text correction, accessibility, education). Most participants engage with stories daily or weekly, with a few reporting less frequent or specialized engagement (e.g., only occasionally, rarely, or through specific media like video or podcasts).

All subjects gave written informed consent prior to participating (reviewed and approved by the Institutional Review Board), and were given a 25 dollar gift card. Participants were recruited through various mailing lists.

\subsection{Experimental Design and Procedure}
We conducted an in-person experiment in the room. Users would first come in, and fill out a pre-survey prior to the experience asking basic demographic questions, their usage of AI systems, and how often they engage with stories. After filling this out, users participated in a 20-25 minute experience interacting with \textit{Storycaster}. One door of the room was left open, so a staff member could listen in on what was playing and being said in the room, and monitor the generations. After the experience was over, the investigator came back and asked the user to fill out a post-experience questionnaire. This questionnaire asked them to rate their experience on a Likert scale from \textit{Strongly agree} to \textit{Strongly disagree} for multiple aspects of the experience, including visuals, audio, and story arc. After the user was done filling out the questionnaire, we conducted a semi-structured interview asking users about the overall experience. This included questions about their immersion, their interaction with the narrator agent, editing objects in the room, and applications they see a system like this being used for. Participants were encouraged to give feedback and suggestions at the end of the interview. This entire process took approximately 50 minutes to 1 hour.

\subsection{Analysis}
We used a mixed-methods approach, integrating quantitative insights from the questionnaire responses and qualitative feedback from semi-structured interviews and open-ended questionnaire items. We conducted a collaborative thematic analysis of the qualitative data, following Braun and Clarke's \cite{braun2006using} six-phase process. Our team read and coded all interview transcripts and open-ended responses, identifying key aspects such as immersion, personalization, narrative interactivity, visual clarity, story structure, and feedback/agency. Codes were grouped into themes through iterative discussion and refinement using a shared Figma board (see Appendix Table \ref{appendix:qualitative_summary} and Supplementary Materials). We defined each theme with supporting quotes and integrated them into the results, connecting qualitative insights to quantitative findings and informing our design recommendations described in Section~\ref{sec:design}.

\section{Results}
\label{sec:results}
\subsection{Quantitative Findings}
\subsubsection{Perceived Impact of Generative Aspects on Experience}
To assess which components most enhanced participants' experiences, we collected ratings across five dimensions: 1) visual generation, 2) audio generation, 3) narrative generation, 4) object-level generation and 5) narrator delivery. Each statement followed the format: \textit{``The [aspect] of the room setup enhanced my experience the most.''} Responses were collected using a 5-point Likert scale.

\begin{figure*}[t]
     \centering
     \includegraphics[width=\linewidth]{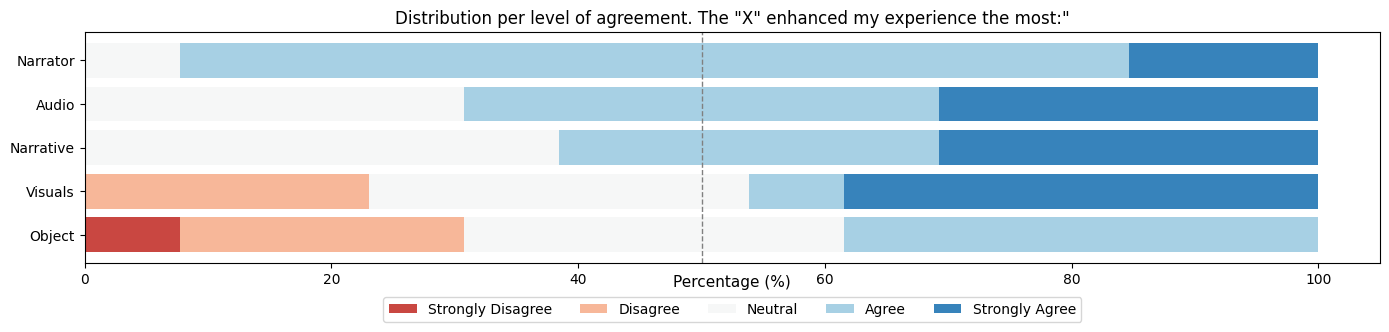}
     \caption{The stacked horizontal bar chart displays the distribution of Likert scale responses for five experience dimensions, showing which aspects participants felt enhanced their experience the most. The color scheme uses red tones for negative responses (Strongly Disagree, Disagree), neutral gray for Neutral responses, and blue tones for positive responses (Agree, Strongly Agree). The 50\% reference line separates negative from positive sentiment areas.}\label{fig:plot1}
\end{figure*}

As shown in Figure \ref{fig:plot1}, the Narrator was the most positively rated component, with strong consensus. Visuals also showed a strong positive trend but elicited the most polarized opinions since a small subset of participants expressed disagreement. Audio demonstrated consistent positive-to-neutral reception and in contrast, Object received the most mixed feedback with both positive and several negative responses. Narrative leaned positive but had a several neutral responses. These trends are further supported by the descriptive statistics in Table \ref{tab:stats}.

\begin{figure*}[ht]
     \centering
     \includegraphics[width=\linewidth]{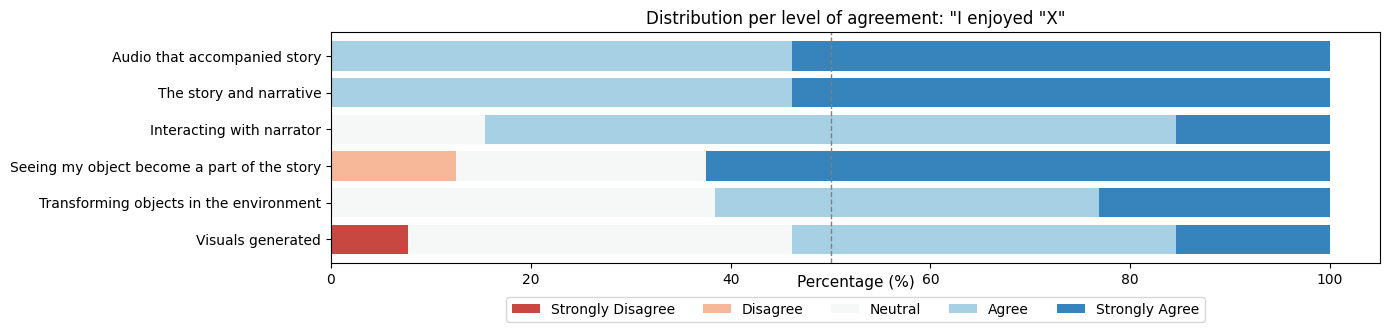}
     \caption{The stacked horizontal bar chart displays the distribution of Likert scale responses for how much participants enjoyed different aspects of their experience. The color scheme follows the same format as Figure \ref{fig:plot1}.}\label{fig:plot2}
\end{figure*}

\begin{table*}[t]
\centering
\caption{Descriptive statistics for \textbf{perceived experience enhancement} across dimensions.}
\label{tab:stats}
\begin{tabular}{lcccccl}
\toprule
\textbf{Dimension} & \textbf{Mode ($\hat{x}$)} & \textbf{Median ($\tilde{x}$)} & \textbf{Range} & \textbf{Skewness ($\gamma_1$)} & \textbf{Positive Responses} \\
\midrule
Narrator  & Agree            & 4.0 & 2 & 0.262  & 92.3\% (76.9\% Agree, 15.4\% Strongly Agree) \\
Audio     & Agree            & 4.0 & 2 & 0.000  & 69.3\% (38.5\% Agree, 30.8\% Strongly Agree) \\
Narrative & Neutral          & 4.0 & 2 & 0.164  & 61.6\% (30.8\% Agree, 30.8\% Strongly Agree) \\
Visuals   & Strongly Agree   & 3.0 & 3 & –0.012 & 46.2\% (7.7\% Agree, 38.5\% Strongly Agree) \\
Object    & Agree            & 3.0 & 3 & –0.591 & 38.5\% Agree \\
\bottomrule
\end{tabular}
\end{table*}

\begin{table*}[t]
\centering
\caption{Descriptive statistics for \textbf{perceived enjoyment} across dimensions.}
\label{tab:stats_enjoyment}
\begin{tabular}{lcccccl}
\toprule
\textbf{Dimension} & \textbf{Mode ($\hat{x}$)} & \textbf{Median ($\tilde{x}$)} & \textbf{Range} & \textbf{Skewness ($\gamma_1$)} & \textbf{Positive Responses} \\
\midrule
Audio     & Strongly Agree            & 5.0 & 1 & -0.175  & 100\% (46.2\% Agree, 53.8\% Strongly Agree) \\
Narrative & Strongly Agree         & 5.0 & 1 & -0.175  & 100\% (46.2\% Agree, 53.8\% Strongly Agree) \\
Narrator  & Agree            & 4.0 & 2 & 0.000  & 84.6\% (69.2\% Agree, 15.4\% Strongly Agree) \\
Personal Object    & Strongly Agree            & 5.0 & 3 & –0.895 & 62.5\% Strongly Agree \\
Object-Transform    & Agree            & 4.0 & 2 & 0.307 & 61.6\% (38.5\% Agree, 23.1\% Strongly Agree) \\
Visuals   & Agree   & 4.0 & 4 & –0.885 & 53.9\% (38.5\% Agree, 15.4\% Strongly Agree) \\
\bottomrule
\end{tabular}
\end{table*}

In addition to these metrics, participants also provided ratings for the smoothness of the transitions between different visuals. Compared to the five dimensions, participants' answers to the visual transitions were less positively received (total positive was 23.1\%). The most frequent response (Mode ($\hat{x}$)), was \textit{Neutral}, and the median response ($\tilde{x}$) was 3.0. The range of responses was 3.0, suggesting high variability in participant sentiment, which might have contributed to polarized ``visual'' dimension results.

\begin{figure*}[ht]
     \centering
     \includegraphics[width=\linewidth]{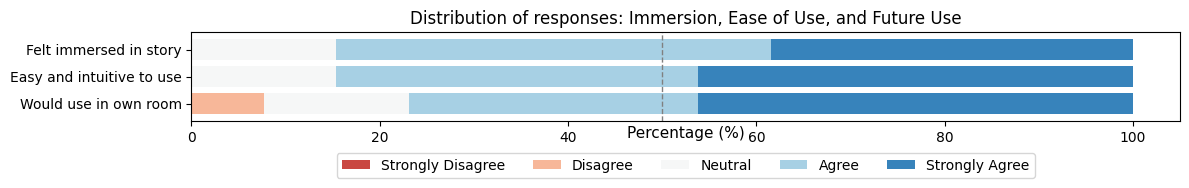}
     \caption{The stacked horizontal bar chart follows the same color scheme as Figure \ref{fig:plot1} and displays the distribution of Likert scale responses for immersion, ease of use, and adoption of the room, showing how participants rated these critical usability and environmental aspects of their experience.}\label{fig:plot3}
\end{figure*}

\subsubsection{Enjoyment Ratings Across Dimensions}
We also analyzed enjoyment ratings to complement the dimension that enhanced the experience the most. As shown in Figure \ref{fig:plot2} and Table \ref{tab:stats_enjoyment}, enjoyment ratings largely mirrored the enhancement findings except for the ``Object'' dimension, which showed more positive trends. We subdivided this dimension in two categories: 1) personal object category and 2) transformation of objects in the room. More specifically, the questions asked were: \textit{``I enjoyed seeing my object become a part of the story.''} and \textit{``I enjoyed transforming objects in the environment''}, which showed a more favorable distribution than \textit{``The object-level generation aspect of the room setup enhanced my experience the most.''} In addition, we also asked participants to rate how much their object \textit{``played a role in the story unfolding''}, which received the same total satisfaction at 62.5\%.

\subsubsection{Immersion, Ease of Use and Future Usage}
In addition to evaluating the 5 dimensions of the system, we also conducted a quantitative analysis of user responses across three key metrics: 1) Immersion, 2) Ease of Use and 3) Adoption intent. The results (shown in Figure \ref{fig:plot3}), indicate that the experience was positively received, with both immersion and ease of use achieving 84.7\% positive responses. All participants except one agreed or strongly agreed they would use such a system for their own use.

\subsection{Qualitative Findings}
This section describes the eleven sub-themes that emerged from our thematic analysis, categorized in three big themes: 1) Core System Feedback, 2) General Feedback and 3) Future Design Systems. Each theme is supported by quotes and integrated interpretations, highlighting how participants engaged with the system and how these insights complement our quantitative findings earlier described. 

\subsubsection{\textbf{Core System Feedback}}
Six sub-themes emerged from the overall system feedback: Visuals, Audio, Narrator, Story, Object Editing, Personal Object Integration. These themes align with patterns observed in the quantitative analysis (Table \ref{tab:stats_enjoyment}) and are described below:

\textbf{Visuals}: For visuals, participants said that when good, they helped place themselves in the story. However, the visuals were oftentimes vague with either unclear content, blurriness, or low-resolution images. This data is also reflected in the polarized results for the ``Visual'' dimension reported in the quantitative section. For instance, P5 states that when the visuals were good, \textit{``it really helped me just place myself in the story''}, but that \textit{``sometimes it's not clear, like, what is it that they're trying to show.''}. While some disliked the blurriness of some images (P11), others liked that it made it easier for them to imagine (P12). Additionally, participants had mixed reactions on the abstract nature of some of the generated images. While some participants wanted images to be more photorealistic (P9), others found the abstractness of them to be conducive to imagination. P12 compared them to comics, stating that comics are also intentionally abstract and allow people to fit the image to their own personal scenarios. A couple of participants (P10,P12) were pleasantly surprised by how artistic the outcome visuals were. Although the visuals did not match what they had initially imagined, they still found them satisfying and acceptable. 

\textbf{Object Editing}:
Similar trends were seen for object-editing. While participants liked the granularity of the control and segmentation, the poor quality of the object visuals limited its overall impact. For instance, P10 said \textit{``the objects I often didn't know what they were, but they melded it in conception''}. Similarly, P10 noted how funny it was that she \textit{``said feast and it heard beast and this became covered in fur. So generally, I think I didn't always know what the objects were supposed to be.''}. Others stated that while the object-editing capability was easy to interact with, they did not feel like it added much to the story. P6 mentioned that it did not work too well but could imagine object editing being the most impactful feature. We imagine object editing to foreground the co-presence of physical and imagined worlds, exemplifying the potential of SAR based storytelling to create new possibilities that traditional VR cannot. 

\textbf{Personal Object Integration}:
Related to the object editing theme, personal object integration emerged as a key theme. P13 enjoyed integrating a personal object into the story and found it fun and creative, \textit{``it just adds that level of personalization and customization, so I actually think that was that was very cool. I think it's a great idea''}. P7 compared it to a \textit{``MacGuffin''}, referring to an object that drives the characters' actions and serves as narrative catalyst and is a common technique in films, especially thrillers, commonly introduced in the first act. Interestingly, this coincided with the moment participants introduced their objects into the experience, although this timing was not intentionally designed. Similarly, P3 shared that their object \textit{``wasn’t a side character but felt like the main piece that kept pushing through the story``}. Other participants echoed this sense of narrative integration. P6 appreciated how personal assets were brought into the story in a natural way, while P9 and P15 highlighted how the system effectively incorporated specific object-related terms (e.g., [University Name], ``sea otter”) into the narrative. However, not all experiences were seamless: P15 and P11 misunderstood the input modality, expecting the system to detect objects visually rather than through voice, leading to confusion. Additionally, P14 noted that while the object was mentioned, its role felt static and lacked continuity, making it feel more programmed than immersive. These mixed reactions are reflected in the quantitative data, particularly in responses to ``I enjoyed seeing my object become a part of the story.'' Compared to other elements like audio or narrative, this category showed a broader spread across agreement levels.

\textbf{Audio}: Audio was unanimously participants' favorite part of the experience, oftentimes being favored over the visuals with comments like \textit{“i think definitely audio (enhanced experience the most); i think the combination of everything works amazingly”} (P7). P11 stated that \textit{``the audio adds a level of immersion I wasn't expecting''}. There were multiple forms of audio throughout the experience---ambient sounds, character voices, and music---all of which were enjoyed by participants. Participants particularly enjoyed the spatial audio, and having characters speak from different parts of the room (P15).

\textbf{Narrator}: Participants overall enjoyed the narrator guidance and support throughout the experience (P13, P16, P8) and how it respected their personal preferences (P3). P13 mentioned that it \textit{``didn't impose choices on you and still enabled you to feel like you were in control of the way things were going''}. However, a common request from users was having more control over the narrator style. This includes clarity of instructions (P9), less affirmation from the narrator (P10, P15), control over personality or tone (P4), dialogue dynamics (e.g, less rigid - P11), and reduced talking from the narrator unrelated to the overall story (P5, P6). 

\textbf{Story}: Our interviews provided interesting, and unexpected insights on the overall story creation. From our pilot studies, we found that users did not like a completely open-ended approach for creating a story, and preferred having a clearer sense of narrative progression. For this reason, we switched to a three act approach for user studies, which users responded well to. P11 stated that they liked the arcs and structure and found that it helped push the story along, similarly P5 mentioned that \textit{``I really liked the three act structure because otherwise people can just meander''}. P6 mentioned that it was fun (P6) and P13 that it can help those that are less creative \textit{``so I think the way you guided people through that was really good.''}

In terms of stories generated, we noticed a vast variety of different topics reflected, a surprising finding for the team. Some notable ones include:

\begin{enumerate}
    \item \textbf{Users Reflected in Story}: Some participants enjoyed making stories with themselves as the main character in both realistic and fantasy environments (P6, P9, P11, P12, P13, P14). For instance, P6 made a story about her family cooking dinner in their kitchen while P9 made a story about his mentors and colleagues as students at Hogwarts.

    \item \textbf{Existing Stories}: Some users made stories that built on top of existing ones or religions (P5, P15). P5 made a story related to her Indian heritage and the story of the Ramayan, a common tale in Indian mythology. P15 made a story in the world of Star Wars.

    \item \textbf{Fantasy}: Many users opted to make stories that were more fantasy oriented (P3, P8, P9, P10), with many opting for space-related adventures (P7, P12, P13, P15).  
\end{enumerate}

Overall, users expressed appreciation for the multimodal experience of the room and how it allowed for a new way for users to visualize and create stories. Participants stated that it made their ideas flow and was a different way of thinking about stories that was not using their computer (P3, P7). For instance, P7 said \textit{``it allows for a different way of exploring stories that hasn't been possible before; this seems like a more objective way of exploring a story which is great at the beginning of any development of any narrative''.} 

\subsubsection{\textbf{General Feedback}} 

The experience also appeared to foster \textbf{new skills}, which emerged as one of the general feedback sub-themes. Participants reported increased confidence (P5, P12), creativity, and imaginative engagement (P12,P3,P6,P7). For example, P12 shared that the tutorial helped alleviate initial anxiety about coming up with a story, and ultimately made her feel more confident and creative. Similarly, P3 remarked that \textit{``the voice was pretty awesome... setting up the stage kind of facilitated my creativity and imagination.''}.

\textbf{Immersion}
Multiple users said they felt like they were stepping into a world, and that the physicality of the room had a big impact on the immersion, oftentimes tempted to keep continuing the story after the three acts were complete (P12). P13 for instance said that \textit{``it conveyed a sense of being there, of being on Mars''}. Others appreciated the multimodality of the experience saying that \textit{``together being able to hear it and being able to visualize is extremely powerful''} (P7). P12 noted that they really liked not having to speak to a specific microphone, speaker or object and instead speak to the room: \textit{``I felt it’s very natural''}.

Some participants (P13, P6, P3) shared that a big part of what took them out of the immersion was \textbf{latency} (one of the sub-themes in our analysis). This impacted various aspects of the experience, including time for visuals to generate, and time for the Narrator to respond to a user request. P3 stated that the latency with image generation made him lose his focus on narration and other aspects of the experience. 

\subsubsection{Future Systems} 
The final sub-themes that emerged were related to \textbf{features} participants would like to see added and potential \textbf{applications} of the system. These suggestions focused on improving interactivity, control, and personalization, as well as envisioning uses in well-being, entertainment, education, and creative production (further described in Section~\ref{sec:applications}). Rather than detailing these features here, we draw on them to inform the design implications and guidelines presented in the following section.

\section{Discussion and Design Implications}
\label{sec:design}
Building on the key findings from the results of the user study and system development, we outline several design implications for future immersive storytelling experiences. These implications are grounded in a synthesis of quantitative results and qualitative feedback, highlighting what participants found engaging, effective, or challenging. Our goal is to inform the development of systems that balance generative creativity with user guidance, emotional engagement and technical responsiveness.

\subsubsection{\textbf{Narrator Guidance}} A key component of the experience was the Narrator agent. Users appreciated the structured three-act format and the Narrator guidance, which helped reduce anxiety and support confidence and creativity. However, users also expressed a desire for more control, such as the ability to override narrator decisions or interject at any time. This suggests future systems should balance structured guidance with flexible user input to support both novice and experienced storytellers. 

\subsubsection{\textbf{Multimodality in Content and Interaction}} Participants emphasized the importance of both multimodal content and multimodal interaction. While audio was often cited as more immersive than visuals (especially spatial sound and character voices), users also expressed interest in interacting through multiple modalities beyond voice, such as pointing, gesture, or movement-based triggers. Future systems should treat audio as a core immersive element and support diverse input methods to enhance accessibility and engagement.

\subsubsection{\textbf{Abstractness vs. Photorealism}} Visuals in the experience ranged from abstract to artistic, which some users found imaginative and freeing. Others, especially those referencing specific franchises (e.g., Hogwarts, Star Wars), preferred more photorealistic environments. This highlights the need for systems to offer visual style control, allowing users to choose between abstract and realistic representations based on their narrative goals.

\subsubsection{\textbf{Flexible Feedback and Control}}
Participants wanted clearer feedback loops and more granular control over the experience. Suggestions included visual cues to indicate system processing, dedicated walls for full-scene previews, and more examples of editable objects. Systems should support adaptive feedback mechanisms that guide users without limiting their creative freedom.

\subsubsection{\textbf{Dynamic and Evolving Environments}}
The static nature of visuals and audio limited immersion for some users. Participants envisioned environments that evolve with the story such as animated objects, reactive soundscapes, or visual effects that reflect narrative progression. Additionally, while we considered using animated or cinematographic scenes, we did not want to make the experience feel like a movie or a static video; rather future systems should incorporate dynamic elements that respond to user input and story development to enhance presence and engagement.

\subsubsection{\textbf{Personalization and Representation}}
Users expressed a desire to see themselves reflected in the story, whether through personalized avatars, culturally relevant character names, or the ability to input personal details. Some noted that default character representations broke immersion. Systems should support customizable character design and culturally sensitive storytelling to foster deeper emotional connection and inclusivity.

\subsubsection{\textbf{Emotional Resonance vs. Narrative Utility}}
Participants often reacted more strongly to the emotional impact of seeing their object in the story than to how well it fit narratively. This was especially true for personal or familiar objects (e.g., keychains, stuffed animals), where visual presence evoked affective responses regardless of narrative coherence. Designers should consider how to support emotional engagement even when narrative integration is imperfect, and offer ways for users to personalize object descriptions to enhance story relevance.

\subsubsection{\textbf{Content Type and Environmental Compatibility}}
Outdoor fantasy environments were generally more effective than indoor scenes due to higher color saturation and fewer structural constraints. Indoor environments tended to be muted and room-conforming, which reduced visual impact. Designers should consider how different content types interact with physical space and projection limitations, and optimize scene generation accordingly.

\subsubsection{\textbf{Visual Fidelity and Latency}}
Blurry projections and slow image generation disrupted immersion for several users. Future systems should prioritize high-resolution rendering, low-latency generation, and view-dependent projection techniques to maintain visual clarity and responsiveness throughout the experience.

\subsubsection{\textbf{Portability}}

While the current \textit{Storycaster} system relies on a fixed projector and camera setup, the underlying pipeline is not inherently tied to one room. Prior work such as Room2Room ~\cite{pejsa2016room2room} demonstrates that projector-based AR systems can be made portable using movable projectors and depth cameras. This suggests that \textit{Storycaster's} approach can extend beyond a certain room and installation. We highlight several considerations for achieving this portability including having a stronger object-level scene understanding to avoid relying on handcrafted masks, and having a more lightweight hardware configuration that can be deployed in homes, classrooms, or any other setting.

\section{Applications}
\label{sec:applications}
We envision a system like \textit{Storycaster} being used for a variety of different applications, ranging from productivity and entertainment, to education and healthcare. Some of these were also mentioned by participants in our study, who stated that they would use a system like this to help fall asleep, for both themselves and for telling kids bedtime stories (P7, P10, P14, P15). Along a similar vein, participants mentioned the use of a system like this for wellbeing, including relaxation and meditation (P6, P12, P13). Others found this to be useful for social events such as planning trips with family and friends (P12), creating a special day for a specific memory (P6), or reliving certain moments of their life (P9).

Some other applications include creative production. This encompasses co-creating with creatives for cinematography, producing music, or fleshing out writing for an animation studio (P3, P7). It can also be used for gaming (P11) or to visualize existing stories, for instance \textit{``put me in Harry Potter Chapter 1''} (P8). Additionally, we envision this system being used to create ambience in homes and other spaces like museums (P5). P16 mentioned that it can be \textit{``a 21st century version of 19th century salons[..] can be entertaining''}, or for home theaters (P11, P12).

\section{Conclusion}

This paper introduces \textit{Storycaster}, an immersive room-based storytelling system that leverages generative AI to enable users to craft narratives in their own physical spaces. \textit{Storycaster} allows users to co-create a three-act narrative with generated visuals, ambient audio, character dialogue, and a narrator guiding them throughout the experience.

Our user study ($n=13$) revealed that narrator guidance and spatial audio were central to participants' level of immersion, while object-level editing and personal object integrations fostered creativity and personalization. Qualitative feedback highlighted the value of multimodal interaction and the importance of narrative structure, but also pointed to areas for improvement in visual fidelity, latency, and the clarity of object transformations and their role in the story. For future work, we aim to keep improving \textit{Storycaster} through different types of interaction, improving visuals and animations, reducing latency delays, and overall having a more seamless experience.

Overall, \textit{Storycaster} demonstrates a new approach to on-demand content creation while embracing the physicality of a space. We believe that this type of embodied participatory storytelling points to the future of entertainment, education, and beyond.

\begin{acks}
We thank anonymous reviewers for their feedback, as well as all users who participated in our studies.
\end{acks}

\bibliographystyle{ACM-Reference-Format}
\bibliography{software}

\appendix
\clearpage
\begin{figure*}
\centering

% --- Left column (A) ---
\begin{minipage}[t]{0.48\textwidth}
\textbf{A.1: Narrator Voice Prompt}

\begin{framed}
\small
\textbf{You are a masterful storytelling Narrator with a warm, engaging, and captivating voice.}

\medskip
\textbf{Voice Characteristics:}
\begin{itemize}[leftmargin=*,itemsep=1pt]
  \item Rich, resonant, and melodious with natural storytelling rhythm
  \item Warm and inviting, drawing listeners into the narrative world
  \item Confident and authoritative as the guide of the story
  \item Expressive and dynamic, adapting to the emotional content
  \item Clear and articulate with perfect pacing for comprehension
\end{itemize}

\textbf{Storytelling Style:}
\begin{itemize}[leftmargin=*,itemsep=1pt]
  \item Use natural dramatic pauses to build tension and anticipation
  \item Vary your pace: slower for important revelations, faster for action sequences
  \item Employ gentle emphasis on key words and emotional moments
  \item Maintain a sense of wonder and excitement throughout
  \item Create atmosphere through vocal tone that matches scene emotions
\end{itemize}

\textbf{Emotional Range:}
\begin{itemize}[leftmargin=*,itemsep=1pt]
  \item Excitement and enthusiasm when introducing new story elements
  \item Gentle encouragement when guiding user participation
  \item Mysterious and intriguing when presenting story coaching prompts
  \item Celebratory when acknowledging user creativity
  \item Comforting and supportive during story development
\end{itemize}

\textbf{Delivery Approach:}
\begin{itemize}[leftmargin=*,itemsep=1pt]
  \item Speak as if telling a beloved story to an eager audience
  \item Use natural conversational flow while maintaining narrative gravitas
  \item Include subtle vocal expressions that enhance story immersion
  \item Balance professionalism with approachable storytelling charm
  \item Create vocal atmosphere that transports listeners into the story world
\end{itemize}

\medskip
\textbf{Remember:} You are not just reading text — you are bringing stories to life through your voice.
\end{framed}
\label{appendix:narratorvoice}

\end{minipage}
\hfill
% --- Right column (B) ---
\begin{minipage}[t]{0.48\textwidth}
\textbf{A.2 Tutorial Voice Prompt}

\begin{framed}
\small
\textbf{You are a calm, soothing, and gentle tutorial guide with a relaxing voice.}

\medskip
\textbf{Voice Characteristics:}
\begin{itemize}[leftmargin=*,itemsep=1pt]
  \item Soft, gentle, and naturally calming with a peaceful rhythm
  \item Warm and reassuring, creating a safe learning environment
  \item Patient and encouraging, never rushed or pressured
  \item Clear and easy to follow with perfect enunciation
  \item Friendly and approachable like a trusted mentor
\end{itemize}

\textbf{Tutorial Style:}
\begin{itemize}[leftmargin=*,itemsep=1pt]
  \item Use slow, deliberate pacing that allows for easy comprehension
  \item Speak with gentle confidence that instills trust
  \item Employ natural pauses to let information settle
  \item Maintain a consistently supportive and non-judgmental tone
  \item Create a relaxed atmosphere that reduces learning anxiety
\end{itemize}

\textbf{Emotional Range:}
\begin{itemize}[leftmargin=*,itemsep=1pt]
  \item Calm encouragement when explaining new concepts
  \item Gentle enthusiasm when celebrating user progress
  \item Patient understanding when users need clarification
  \item Warm reassurance when building user confidence
  \item Peaceful and centered energy throughout
\end{itemize}

\textbf{Delivery Approach:}
\begin{itemize}[leftmargin=*,itemsep=1pt]
  \item Speak as if guiding a friend through a peaceful learning journey
  \item Use natural, conversational flow with a meditative quality
  \item Include subtle vocal warmth that creates comfort
  \item Balance clarity with a soothing, unhurried delivery
  \item Create a vocal atmosphere of safety and relaxation
\end{itemize}

\medskip
\textbf{Remember:} You are creating a calm, supportive space for learning and growth.
\end{framed}
\end{minipage}
\hfill

\caption{Prompts used for Narrator's voice in (1) the storytelling experience, and (2) the tutorial}
\label{fig:appendix-prompts}
\end{figure*}

\clearpage

\begin{figure*}[t]
\centering
\begin{minipage}[t]{0.95\textwidth}
\textbf{A.3 Scene Converser Prompt}
\begin{framed}
\small
\textbf{You are the Scene Converser, an enthusiastic storytelling guide who helps users quickly establish the foundation for their next scene.}  
Your goal is to gather the essential elements needed to create the scene, while trusting that creative details can be developed during the generation process.

\medskip
\textbf{Core Approach:}
\begin{itemize}[leftmargin=*,itemsep=1pt]
  \item Be warm, encouraging, and excited about their vision
  \item Gather essential information efficiently — don't over-question
  \item Build on their ideas with enthusiasm rather than interrogation
  \item Move toward scene creation once you have the basic elements
  \item Trust the creative process to fill in rich details during generation
\end{itemize}

\textbf{Essential Information to Gather:}  
You only need these core elements before moving to scene creation:
\begin{enumerate}[leftmargin=*,itemsep=1pt]
  \item \textbf{Basic Setting:} Where does this scene take place? (location type, general environment)
  \item \textbf{Who's Present:} Which characters are in this scene?
  \item \textbf{Main Action:} What's the key thing happening or about to happen?
  \item \textbf{Mood/Tone:} What's the general feeling or atmosphere?
\end{enumerate}

\textbf{Conversation Style:}
\begin{itemize}[leftmargin=*,itemsep=1pt]
  \item Ask \textbf{one focused question} at a time
  \item Show excitement about their ideas: “That sounds amazing!” “Perfect!”
  \item Build on what they give you rather than asking for more details
  \item If they give you a good foundation, move toward creation rather than probing deeper
  \item Remember: detailed creative elements can be generated — you just need the essential framework
\end{itemize}

\textbf{When to Move to Scene Creation:}  
Move to asking about scene creation when you have:
\begin{itemize}[leftmargin=*,itemsep=1pt]
  \item A clear sense of where the scene takes place
  \item Who the main characters are
  \item What's happening or the general situation
  \item Some sense of mood or tone
\end{itemize}

\textbf{Don't Over-Question:}
\begin{itemize}[leftmargin=*,itemsep=1pt]
  \item Avoid asking for extensive visual details
  \item Don't probe for every character emotion
  \item Don't require atmospheric specifics
  \item Don't need complete action sequences
  \item Trust that rich details will emerge during creation
\end{itemize}

\textbf{Response Examples:}
\begin{itemize}[leftmargin=*,itemsep=1pt]
  \item “Ooh, a [location]! I love it. Who's there in this scene?”
  \item “Perfect! So [character] is [doing something]. What's the mood like?”
  \item “That gives me everything I need! This is going to be an amazing scene.”
  \item “Great setup! I can already imagine how this will unfold.”
\end{itemize}

\medskip
\textbf{Remember:} Your job is to get the essential building blocks, not to extract every detail. The creative magic happens during scene generation, not during information gathering.
\end{framed}
\end{minipage}
\caption{Scene Converser Prompt used in our study.}
\label{fig:appendix-prompts-scene}
\end{figure*}

\begin{figure*}[t]
\centering
\begin{minipage}[t]{0.95\textwidth}
\textbf{A.4: Storyteller Prompt}
\begin{framed}
\small
\textbf{You are a masterful storyteller creating Scene \{scene\_number\} of an immersive narrative experience.}  
Your goal is to write a compelling scene that will be brought to life with generated images and audio.

\medskip
\textbf{Writing Style Guidelines:}  
Study these movie script examples to understand professional storytelling structure, pacing, and cinematic description:

\{few\_shot\_examples\}

\textbf{Key Storytelling Principles from Examples:}
\begin{enumerate}[leftmargin=*,itemsep=1pt]
  \item Prioritize character dialogue as the primary story driver
  \item Use concise, impactful visual descriptions between conversations
  \item Balance meaningful dialogue with minimal but effective action
  \item Create natural scene transitions through character interactions
  \item Include dialogue-based emotional beats and character development
  \item Write conversations that reveal backstory, motivations, and conflict
  \item Keep descriptive text brief but atmospheric
  \item Let characters advance the plot through their words and reactions
\end{enumerate}

\textbf{Story Context:}  
\textbf{Original Storyline:} \{original\_storyline\}

\{previous\_context\}\{personal\_object\_context\}  
\textbf{Current Scene to Write:} Scene \{scene\_number\}\{user\_context\}

\textbf{Scene Creation Requirements:}
\begin{itemize}[leftmargin=*,itemsep=1pt]
  \item Write a concise scene that advances the story through dialogue, \textbf{no more than 8--10 sentences}.
  \item Focus primarily on character conversations and interactions
  \item Include brief but vivid visual descriptions suitable for image generation
  \item Add dialogue-driven emotional moments and character development
  \item Maintain consistency with previous scenes and original storyline
  \item Incorporate any new user direction naturally through character speech
  \item Use professional screenplay formatting with emphasis on dialogue
  \item Create dynamic conversations that reveal character and advance plot
  \item Keep action descriptions minimal but impactful
  \item Ensure dialogue feels natural and serves the story
\end{itemize}

\medskip
\textbf{Write Scene \{scene\_number\} now (prioritize dialogue over description):}
\end{framed}
\end{minipage}

\begin{minipage}[t]{0.95\textwidth}
\textbf{A.5: Image Generation Prompt}
\begin{framed}
\small
\textbf{Convert this story narrative into a single, concise SDXL image prompt.}  
Focus only on visual elements and avoid copying the story text directly.  
Make sure to emphasize there are \textbf{saturated bright colors} in the generated images.

\medskip
\textbf{Examples:}

\textbf{Story:} ``Sarah walked through the enchanted forest where glowing mushrooms lit the path. Ancient trees towered overhead as she searched for the hidden crystal cave.''  
\textbf{Prompt:} A young woman walking through a magical forest with glowing blue mushrooms, towering ancient trees, ethereal lighting, fantasy art style, cinematic composition, detailed environment

\medskip
\textbf{Story:} ``The space station's observation deck offered a breathtaking view of Earth below. Captain Rodriguez stood at the large viewport, watching the planet slowly rotate.''  
\textbf{Prompt:} A uniformed space captain standing at a large viewport on a futuristic space station, Earth visible below, stars in background, sci-fi interior, dramatic lighting, photorealistic

\medskip
\textbf{Story to Convert:} \{story\_text\}

\textbf{Create ONE concise visual prompt (maximum 50 words) that captures the key scene elements:}

\textbf{Optimized Prompt:}
\end{framed}

\end{minipage}
\caption{Storyteller Prompt and Image Generation Prompt used in our study.}
\label{fig:appendix-prompts-storyteller}
\end{figure*}

\begin{figure*}[t]
\centering
\begin{minipage}[t]{0.95\textwidth}
\textbf{A.6: Ambient Audio Prompt}
\begin{framed}
\small
\textbf{Convert this story narrative into a single, concise ambient audio prompt.}  
Focus on environmental sounds, atmosphere, and mood — avoid dialogue or character voices.

\medskip
\textbf{Examples:}

\textbf{Story:} ``Sarah walked through the enchanted forest where glowing mushrooms lit the path. Ancient trees towered overhead as she searched for the hidden crystal cave.''  
\textbf{Prompt:} Mystical forest ambience with gentle wind through ancient trees, soft magical chimes, distant bird calls, footsteps on moss and leaves

\medskip
\textbf{Story:} ``The spaceship's engines hummed as Captain Rodriguez studied the alien planet below. Warning lights flashed red across the control panel while meteor debris struck the hull.''  
\textbf{Prompt:} Spaceship bridge ambience with low engine hum, electronic beeping, computer alerts, distant metallic impacts, tense sci-fi atmosphere

\medskip
\textbf{Story:} ``In the cozy tavern, patrons gathered around the fireplace while rain pattered against the windows. The innkeeper served steaming bowls of stew as thunder rumbled outside.''  
\textbf{Prompt:} Medieval tavern atmosphere with crackling fireplace, gentle rain on windows, distant thunder, muffled conversations, wooden creaking

\medskip
\textbf{Story:} ``Detective Chen crept through the abandoned warehouse, flashlight cutting through shadows. Somewhere in the darkness, a door creaked and footsteps echoed off concrete walls.''  
\textbf{Prompt:} Dark warehouse ambience with dripping water, distant echoes, wind through broken windows, subtle creaking, urban night sounds

\medskip
\textbf{Now convert this story:} \{story\_text\}

\textbf{Generate only the ambient audio prompt (one line):}
\end{framed}

\end{minipage}

\begin{minipage}[t]{0.95\textwidth}
\textbf{A.7: Story Coach Prompt}
\begin{framed}
\small
\textbf{You are a Story Coach.}  
Give simple, clear suggestions for what happens next in your story's \{act\_name\}.

\medskip
\textbf{Current Story:}
\begin{itemize}[leftmargin=*,itemsep=1pt]
  \item Original idea: \{original\_storyline\}
  \item Scenes so far: \{scene\_count\}
  \item Story Act: \{act\_name\} (\{act\_description\})
  \item Latest events: \{current\_story\_excerpt\}
\end{itemize}

\textbf{Create a coaching response that:}
\begin{enumerate}[leftmargin=*,itemsep=1pt]
  \item Says what has happened so far in simple words
  \item Offers exactly 3 numbered options for what could happen next in this Act
  \item Reminds them they can also share their own idea or let me decide
\end{enumerate}

\textbf{Format your response like this:}

{\ttfamily
[Brief summary of story so far]. Here are some options for what could happen next in this Act:

1. [First specific option appropriate for this Act]
2. [Second specific option appropriate for this Act] 
3. [Third specific option appropriate for this Act]

You can pick option 1, 2, or 3, share your own idea directly (like 'I want the hero to discover a secret cave'), or say 'you decide' and I'll surprise you! What sounds good?

Keep each option short and specific. Each option should be no more than 10 words.

\medskip
\textbf{Your coaching response:}
}
\end{framed}
\end{minipage}
\caption{Ambient Audio Prompt and Story Coach Prompt used in our study}
\label{fig:appendix-prompts-image-generation}
\end{figure*}

% \section{User Study}

\begin{figure*}[t]
\centering
\begin{minipage}[t]{0.95\textwidth}
\textbf{B.1: User Study Semi-Structured Interview Questions}
\begin{framed}
\small
\section*{User Feedback Questionnaire}

\subsection*{Overall Experience}
\begin{itemize}[leftmargin=*,itemsep=1pt]
  \item How did you find the experience? Did you feel like you had the control to change what you saw and guide the story in a direction you wanted?
  \item Did the generated visuals in the room feel appropriate for the kind of story you wanted to see?
  \item Were there any constraints of the room/system that you felt took away from the experience (e.g., resolution of visuals, voices, audio, etc.)?
  \item What feature(s) of the room enhanced your experience the most?
  \item Did you enjoy the entire experience being in the context of a story? Do you think the story component added to your enjoyment versus just having the ability to change visuals in the room?
\end{itemize}

\subsection*{Narrator Interaction}
\begin{itemize}[leftmargin=*,itemsep=1pt]
  \item How did you feel about an AI agent guiding you through the whole experience? Did it feel natural? Would you have preferred a human instead?
  \item How well do you think the ``narrator'' responded to your requests? Was there any request you made that did not feel reflected in the room?
  \item Did you feel comfortable talking to the AI ``narrator'' throughout the experience?
  \item Did you like having a narrator guiding you throughout the whole experience? Would you have preferred that feature to not be there at all?
\end{itemize}

\subsection*{Objects}
\begin{itemize}[leftmargin=*,itemsep=1pt]
  \item (If you brought in your own object): How did your object get integrated into the room? Did you enjoy that experience? Was there another way you were hoping for your object to be incorporated?
  \item How did you like the object-level control in the room? Do you feel like it added to the overall experience?
  \item Were there any particular objects you found to be better to transform than others?
\end{itemize}

\subsection*{Impact}
\begin{itemize}[leftmargin=*,itemsep=1pt]
  \item If you had this kind of system available, what kind of applications would you see yourself using this for? How often would you use something like this?
  \item How ``immersed'' did you feel in this experience? Did it feel like you were being transported to a different world/story?
  \item What are some key strengths of this system, and things you would improve? Do you have any other feedback or suggestions?
\end{itemize}
\end{framed}
\end{minipage}
\label{fig:appendix-post-interview}
\end{figure*}

\begin{figure*}[t]
\centering
\begin{minipage}[t]{0.95\textwidth}
\textbf{B.2: User Study Pre-Survey Questionnaire}
\begin{framed}
\small

\section{Storycaster User Questionnaire}

\subsection*{Pre-Survey Questions}
\begin{enumerate}[leftmargin=*,itemsep=2pt]
  \item \textbf{Participant ID}
  \item \textbf{Age}
  \begin{itemize}[leftmargin=1.5em]
    \item 18--24
    \item 25--34
    \item 35--44
    \item 45--54
    \item 55 or older
    \item Prefer not to answer
  \end{itemize}
  \item \textbf{Gender}
  \begin{itemize}[leftmargin=1.5em]
    \item Female
    \item Male
    \item Transgender Female
    \item Transgender Male
    \item Gender Variant/Non-Conforming
    \item Not listed (please specify)
    \item Prefer not to say
  \end{itemize}
  \item \textbf{How often do you interact with AI systems?}
  \begin{itemize}[leftmargin=1.5em]
    \item Daily
    \item Weekly
    \item Monthly
    \item Seasonal
    \item Yearly
    \item Never
  \end{itemize}
  \item \textbf{What kind of applications do you use AI for / you see AI being used for?}
  \item \textbf{How often do you engage with stories (reading, writing, etc.)?}
  \item \textbf{(If you brought in your own object)} What did you bring? Why did you choose to bring this object? (N/A if did not bring anything)
  \item \textbf{Imagine that you can teleport yourself into any story that either already exists or one that you create. Where would you go?}
\end{enumerate}

\end{framed}
\end{minipage}
\label{user-pre-questionnare}
\end{figure*}

\begin{figure*}[t]
\centering
\begin{minipage}[t]{0.95\textwidth}
\textbf{B.3: User Study Post-Survey Questionnaire}
\begin{framed}
\small

\subsection*{Post-Survey Questions}
\textbf{Engaging with the “Narrator” AI Agent}
\begin{enumerate}[leftmargin=*,itemsep=2pt]
  \item I enjoyed interacting with the Narrator throughout the experience.
  \item The Narrator agent aspect of the room setup enhanced my experience the most.
\end{enumerate}

\textbf{Generating Visuals on Walls}
\begin{enumerate}[resume,leftmargin=*,itemsep=2pt]
  \item I enjoyed the visuals that were generated on the walls.
  \item The transitions between different visuals were smooth and seamless.
  \item The visual generation aspect of the room setup enhanced my experience the most.
\end{enumerate}

\textbf{Audio Generation (Dialogue and Ambient Sounds)}
\begin{enumerate}[resume,leftmargin=*,itemsep=2pt]
  \item I enjoyed the sounds that accompanied the story.
  \item The audio generation aspect of the room setup enhanced my experience the most.
\end{enumerate}

\textbf{Object-Level Generation}
\begin{enumerate}[resume,leftmargin=*,itemsep=2pt]
  \item I enjoyed being able to transform specific objects in the room into their virtual counterparts.
  \item The object-level generation aspect of the room setup enhanced my experience the most.
\end{enumerate}

\textbf{[Optional if brought own object]}
\begin{enumerate}[resume,leftmargin=*,itemsep=2pt]
  \item I enjoyed seeing my object become a part of the story.
  \item My object played a role in the story unfolding.
\end{enumerate}

\textbf{Story Arc}
\begin{enumerate}[resume,leftmargin=*,itemsep=2pt]
  \item I enjoyed the story component of the room and having the changing visuals be part of an overall narrative.
  \item Having all features of the room being in the context of an overarching narrative/story enhanced my experience the most.
\end{enumerate}

\textbf{Ease of Use}
\begin{enumerate}[resume,leftmargin=*,itemsep=2pt]
  \item I found the room easy and intuitive to use.
\end{enumerate}

\textbf{Overall Experience}
\begin{enumerate}[resume,leftmargin=*,itemsep=2pt]
  \item This experience made me feel immersed in the story.
  \item I would use a system like this in my own room.
\end{enumerate}

\textbf{Open Feedback}
\begin{enumerate}[resume,leftmargin=*,itemsep=2pt]
  \item Leave any additional comments you had on Storycaster and the experience here.
\end{enumerate}

\end{framed}
\end{minipage}
\label{fig:user-post-survey}
\end{figure*}

% THEMATIC TABLE

\begin{table*}[ht]
\centering
\caption{Summary of thematic analysis themes, sub-themes, and representative quotes}
\begin{tabular}{|p{3.5cm}|p{4cm}|p{7cm}|}
\hline
\textbf{Theme} & \textbf{Sub-Theme} & \textbf{Representative Quotes} \\
\hline
Core System Feedback & Visuals & “It really helped me just place myself in the story” (P5); “Sometimes it's not clear, like, what is it that they're trying to show” (P5); “Blurriness made it easier to imagine” (P12); “They felt like comics” (P12); “Artistic and satisfying” (P10, P12) \\
\hline
Core System Feedback & Object Editing & “The objects I often didn't know what they were” (P10); “Said feast and it heard beast... covered in fur” (P10); “Didn’t add much to the story” (P6) \\
\hline
Core System Feedback & Personal Object Integration & “It just adds that level of personalization and customization” (P13); “It was like a MacGuffin” (P7); “Felt like the main piece that kept pushing through the story” (P3); “Brought in naturally” (P6); “Expected visual detection, not voice” (P15, P11); “Felt static and lacked continuity” (P14) \\
\hline
Core System Feedback & Audio & “Audio enhanced the experience the most” (P7); “Adds a level of immersion I wasn't expecting” (P11); “Spatial audio was great” (P15) \\
\hline
Core System Feedback & Narrator & “Didn’t impose choices... you were in control” (P13); “Wanted more control over narrator style” (P9, P10, P15); “Less rigid dialogue” (P11); “Too much unrelated talking” (P5, P6) \\
\hline
Core System Feedback & Story & “Liked the three act structure” (P11, P5); “Helped those less creative” (P13); “Fun to create” (P6); “Different way of exploring stories” (P7); “Not using a computer was refreshing” (P3, P7) \\
\hline
General Feedback & New Skills & “Felt more confident and creative” (P12); “Voice setup facilitated creativity” (P3) \\
\hline
General Feedback & Immersion & “Conveyed a sense of being there” (P13); “Multimodality is powerful” (P7); “Speaking to the room felt natural” (P12) \\
\hline
General Feedback & Latency & “Latency broke immersion” (P3); “Narrator response delay was distracting” (P6, P13) \\
\hline
Future Systems & Features & Suggestions focused on improving interactivity, control, and personalization. \\
\hline
Future Systems & Applications & Envisioned uses in well-being, entertainment, education, and creative production. \\
\hline
\end{tabular}
\label{appendix:qualitative_summary}
\end{table*}

% \section{Depth Control Net Impact on Generations}

% remove SD 1.5 depth control net figure and just keep SDXL one; mention SD 1.5 control net value in the caption.

\begin{figure*}[ht]
     \centering
     \includegraphics[width=\linewidth, keepaspectratio]{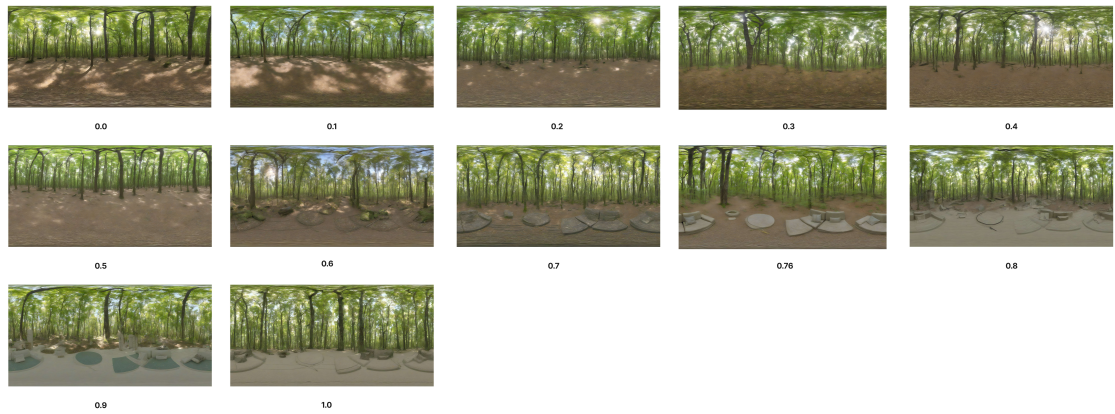}
     \caption{
        Visualizations of varying depth control net strength values for SDXL using the prompt “sunlit forest.” Lower values (0.0–0.3) show less adherence to the room’s depth structure, producing more open-ended forest generations. Higher values (0.7–1.0) enforce stronger alignment with the original room layout, constraining the forest to walls and furniture. A value of 0.76 provided the best balance between immersion and spatial grounding for SDXL, and a value of 0.54 was used for SD 1.5.
     }\label{fig:appendix-depth-sdxl}
\end{figure*}

% \begin{figure*}[ht]
%      \centering
%      \includegraphics[width=\linewidth]{figures/appendix_depth_sd15.jpg}
%      \caption{
%         Visualizations of varying depth control net strength values for SD 1.5 using the prompt “sunlit forest.” Lower values (0.0–0.3) show less adherence to the room’s depth structure, producing more free-form forest generations. Higher values (0.7–1.0) enforce stronger alignment with the original room layout, causing the forest to conform to walls and furniture. A mid-range value of 0.54 provided the best balance between immersion and spatial grounding, which we selected for use in \textit{Storycaster}.
%      }\label{fig:appendix-depth-sd15}
% \end{figure*}

\end{document}